\documentclass[12pt]{amsart}

\newcommand{\kms}{\mbox{km s$^{-1}$}}

\newcommand{\Msun}{\mbox{M$_{\sun}$}}

\newcommand\arcdeg{\mbox{$^\circ$}}%

\newcommand\farcs{\mbox{$.\!\!^{\prime\prime}$}}
\newcommand\ion[2]{#1$\;${\small\rmfamily\@Roman{#2}}\relax}%
\newcommand\degr{\arcdeg}%
\newcommand\sun{\odot}%
\usepackage[letterpaper]{geometry}
\usepackage{setspace}
\usepackage[super]{natbib}
\usepackage{amsmath}
\usepackage{amssymb}
\usepackage{graphicx}
\usepackage{lipsum}

\makeatletter
\newcommand{\authorfootnotes}{\renewcommand\thefootnote{\@fnsymbol\c@footnote}}%
\makeatother
\begin{document}

\title{A Triple Protostar System Formed via Fragmentation of a Gravitationally Unstable Disk}

\maketitle
\authorfootnotes
\begin{center}
  John J. Tobin\textsuperscript{1,2}, Kaitlin M. Kratter\textsuperscript{3}, Magnus V. Persson\textsuperscript{2,4},
  Leslie W. Looney\textsuperscript{5}, Michael M. Dunham\textsuperscript{6}, Dominique Segura-Cox\textsuperscript{5},
  Zhi-Yun Li\textsuperscript{7}, Claire J. Chandler\textsuperscript{8}, Sarah I. Sadavoy\textsuperscript{9},
  Robert J. Harris\textsuperscript{5}, Carl Melis\textsuperscript{10}, Laura M. P{\'e}rez\textsuperscript{11}
   \par \bigskip
\end{center}

\noindent \textsuperscript{1}{Homer L. Dodge Department of Physics and Astronomy, University of Oklahoma, 440 W. Brooks Street, Norman, OK 73019, USA}\\
\textsuperscript{2}{Leiden Observatory, Leiden University, P.O. Box 9513, 2300-RA Leiden, The Netherlands}\\
\textsuperscript{3}{Department of Astronomy and Steward Observatory, Univ.  of Arizona, 933 N Cherry Ave, Tucson, AZ, 85721 USA}\\
\textsuperscript{4}{Department of Earth and Space Sciences, Chalmers University of Technology, Onsala Space Observatory, 439 92, Onsala, Sweden}
\textsuperscript{5}{Department of Astronomy, University of Illinois, Urbana, IL 61801, USA}\\
\textsuperscript{6}{Department of Physics, SUNY Fredonia, Fredonia, New York 14063, USA}\\
\textsuperscript{7}{Department of Astronomy, University of Virginia, Charlottesville, VA 22903, USA}\\
\textsuperscript{8}{National Radio Astronomy Observatory, P.O. Box O, Socorro, NM 87801, USA}\\
\textsuperscript{9}{Max-Planck-Institut f\"ur Astronomie, D-69117 Heidelberg, Germany}\\
\textsuperscript{10}{Center for Astrophysics and Space Sciences, University of California, San Diego, CA 92093, USA}\\
\textsuperscript{11}{Max-Planck-Institut f\"ur Radio Astronomie, Auf dem H\"ugel 69, 53121, Bonn, Germany}\\

\textbf{Binary and multiple star systems are a frequent outcome of the star formation
process\citep{duchene2013,reipurth2014}, and as a result, almost 
half of all sun-like stars have at least one companion star\citep{raghavan2010}. 
Theoretical studies indicate that there are two main pathways 
that can operate concurrently to form binary/multiple star systems: 
large scale fragmentation of turbulent gas cores and filaments\cite{fisher2004,padoan2004} 
or smaller scale fragmentation of a massive protostellar disk due to gravitational 
instability\citep{adams1989,bonnell1994b}. 
Observational evidence for turbulent fragmentation on scales 
of $>$1000~AU has recently emerged\citep{pineda2015,lee2016}.
Previous evidence for disk fragmentation was limited
to inferences based on the separations of more-evolved pre-main sequence and protostellar multiple systems\citep{connelley2008,kraus2011,takakuwa2012,tobin2016}. The triple protostar 
system L1448 IRS3B is an ideal candidate to search for evidence of disk 
fragmentation. L1448 IRS3B is in an early phase of the star formation process, 
likely less than 150,000 years in age\citep{lee2015}, and all protostars in
the system are separated by $<$200~AU.
Here we report observations of dust and molecular gas emission that
reveal a disk with spiral structure surrounding the three protostars.
Two protostars near the center of the disk are separated by 61 AU, and a tertiary
protostar is coincident with a spiral arm in the outer disk at a 183 AU separation\citep{tobin2016}. 
The inferred mass of the central pair of protostellar objects is $\sim$1 \Msun, 
while the disk surrounding the three protostars has a total 
mass of $\sim$0.30 M$_{\sun}$. The tertiary protostar itself has a minimum mass of $\sim$0.085 M${\sun}$.
We demonstrate that the disk around L1448 IRS3B appears susceptible to disk fragmentation
at radii between 150~AU and 320~AU, overlapping with the location of the tertiary protostar. 
This is consistent with models for a protostellar disk that has recently undergone 
gravitational instability, spawning one or two companion stars.}


L1448 IRS3B is located in the Perseus molecular cloud at a distance of $\sim$230~pc\citep{hirota2011}
and contains three protostars out of the six that collectively make up 
L1448 IRS3\citep{tobin2016,lee2015}, spanning 0.05 pc.
L1448 IRS3B is a Class 0 protostar system\citep{sadavoy2014}, which signifies an
early phase of the star formation process when the protostars
are deeply enshrouded in an envelope of accreting material\citep{andre1993}. 
The three protostars in L1448 IRS3B (denoted -a, -b, and -c) have a hierarchical configuration; 
the central-most protostar, IRS3B-a, has projected separations
from IRS3B-b and IRS3B-c of 61~AU and 183~AU, 
respectively\citep{tobin2016}. The new observations of L1448 IRS3B conducted with the 
Atacama Large Millimeter/submillimeter Array (ALMA) at a resolution of 
0\farcs27$\times$0\farcs16 (62~AU~$\times$~37~AU)
provide images at 1.3~mm of the dust and gas emission surrounding the three protostars with 10$\times$
higher sensitivity and 2$\times$ higher resolution than previous studies.

The ALMA 1.3~mm image of L1448 IRS3B is shown in Figure 1, revealing dust emission toward
each of the three distinct protostars identified in previous Karl G. Jansky Very
Large Array (VLA) observations\citep{tobin2016}. The ALMA images also reveal a disk with substructure  
surrounding the entire system, extending to a radius of $\sim$400~AU.
The disk appears to have a dominant one or two-armed spiral that links
IRS3B-a and IRS3B-b with the more widely separated
IRS3B-c, which is embedded in the outermost arm. The disk geometry and rotation profile (see below) 
place the system center of mass near the close pair, IRS3B-a and IRS3B-b, rather than toward the
more distant (and brighter at 1.3~mm and 8~mm) IRS3B-c. 
A VLA 8~mm image is also shown in Figure 1, having comparable resolution to the ALMA data.
The 8~mm emission toward the protostar positions 
may be a combination of both concentrated dust emission and free-free emission from the base of the
jets driven by the protostars\citep{anglada1998}. However, the weak 8~mm emission associated
with the disk and spiral arms surrounding the three protostars is from dust only.
The weaker emission from the disk at 8~mm is unsurprising, because thermal dust 
emission at sub-millimeter and longer wavelengths decreases in brightness with the 
inverse square of wavelength, modified by the dust opacity spectral index 
(see Methods Section 3).

The resolved disk in the ALMA observations also enables us to estimate the 
inclination of the system to be 45.4\degr, allowing us to calculate the deprojected separations of the 
companions (see Methods section 2). The deprojected separations of IRS3B-b 
and IRS3B-c from IRS3B-a are 78~AU and 254~AU, respectively.
We estimate the mass of the entire disk using the dust continuum emission (with IRS3B-c removed). 
Assuming the dust and gas are well mixed, we find $M_{\rm disk}=0.30~M_{\sun}$ using
the integrated 1.3~mm flux density of 0.52~$\pm$~0.002~Jy. Finally, the mass of IRS3B-c in the outer disk 
is estimated to be 0.085~$M_{\sun}$ from the 1.3~mm flux density of 
0.19~$\pm$~0.003~Jy; this is a lower limit to the gas mass given that IRS3B-c 
is likely opaque at 1.3~mm (see Methods Section 3). The surface density profile
of the disk is subsequently derived from the 1.3~mm continuum data (see Methods Section 4).

The velocity structure of the disk is shown in Figure 2, using the Doppler shift of the C$^{18}$O 
molecular line emission as the gas orbits the center of mass in the system. The distinct
separation between the blue and red-shifted emission is indicative of a rotation
pattern centered on IRS3B-a and IRS3B-b. The rotation pattern reinforces the conclusion from the
dust continuum that IRS3B-a and IRS3B-b comprise the dominant mass in the system. Furthermore, 
the overall mass of IRS3B-c must be low relative to IRS3B-a and IRS3B-b, despite it being
the brightest source at 1.3~mm and 8~mm.
We estimate that IRS3B-a and IRS3B-b have a combined  mass of $\sim$1.0~$M_{\sun}$ from the disk rotation profile (see Methods Section 5). We are unable to make an estimate
of the mass for IRS3B-c from molecular line kinematics, but the 0.085~$M_{\sun}$ from the dust emission
can be considered a lower limit. Note that the C$^{18}$O emission does not trace 
the entire disk structure seen in the dust because any emission at velocities
within $\pm$$\sim$1~\kms\ of the system velocity ($\sim$4.5~\kms) is 
filtered-out along with the large, surrounding envelope.  

With these physical parameters in hand, we can now assess whether 
L1448 IRS3B hosts a gravitationally unstable disk that recently
fragmented to form IRS3B-c. Gravitational instabilities arise in 
protostellar disks when the self-gravity of the disk becomes 
significant compared to the gravity of the central protostar(s). 
While disk fragmentation is the non-linear outcome of a complex hydrodynamic instability,
the canonical analytic estimate for when instability 
occurs is that the Toomre $Q$ parameter\citep{toomre1964} is of order 
unity (see the Methods section 6 for details). For Keplerian disks, Toomre's $Q$ can be expressed as:
\begin{equation}
\label{eq:qapprox}
Q \approx 2\frac{M_*}{M_d}\frac{H}{r}
\end{equation}
where $H=c_s/\Omega$ is the disk scale height, $c_s$ the disk sound speed, and $\Omega$ the Keplerian angular velocity. Using the masses derived above, we can immediately see that for
typical disk temperatures of $30$~K ($c_s$~=~0.3~\kms) at a 
radius of 200~AU ($\Omega$~$\sim$~7$\times$10$^{-11}$~s$^{-1}$),  
$Q\approx 1$. While suggestive of instability, a more robust model is desirable.
In the Methods Section 6, we describe a simple analytic disk model\citep{kratter2016} illustrating 
that the IRS3B disk likely underwent gravitational instability in the recent past.
Our analytic model is consistent with many previous numerical simulations of fragmenting disks 
that replicate this type of disk morphology (e.g., spiral arms) for parameters comparable
to the observed quantities\citep{kratter2010,bate2012}.

We show the result of our analysis in Figure 3, where we plot the range of possible $Q$ values for
different inferred mass accretion rates, as a function of disk radius. Given the mass of the central 
protostars (both assumed to be
$\sim$0.5~M$_{\sun}$) and the luminosity of 3.2~L$_{\sun}$\citep{murillo2016}, a current mass accretion
rate of $\sim$10$^{-7}$ M$_{\sun}$~yr$^{-1}$ is most consistent with the model. 
For a wide range of accretion rates, the disk around  
IRS3B is marginally unstable ($Q\approx1$) at 
radii between 150~AU and 320~AU. This conclusion is consistent with both 
the continued presence of the spiral arms and IRS3B-c forming near its current deprojected
separation of 254~AU. The evolutionary phase of IRS3B-c is also consistent with it 
forming within the last few dynamical times ($\sim$4000 yr assuming Keplerian rotation around IRS3B-a/b).
Prior to the formation of IRS3B-c, the outer disk is likely to have been more unstable, 
and the subsequent accretion of disk mass onto IRS3B-c is now stabilizing the disk.
A fragment formed by instability 
in this disk would likely begin with a mass of roughly $10^{-2}$~$M_\odot$, 
consistent with IRS3B-c undergoing accretion to 
arrive at our measured mass of at least $0.085M_\odot$ (see Methods Section 6).

In addition to IRS3B-c, IRS3B-b may have formed from an earlier episode of disk fragmentation\citep{kratter2010}. IRS3B-b could have formed near its current location, if the disk was 
more unstable in the past. It could also have formed in the outer disk prior to IRS3B-c and migrated inward
to its current location over a few 10s of orbital timescales\citep{zhu2012}.
Fragments formed early in the evolution of the disk can grow to comparable mass 
as the primary star\citep{kratter2010}. 
Finally, we note that if all three protostars were on roughly circular, co-planar, Keplerian 
orbits with semi-major axes equal to their deprojected separations, they would be marginally 
dynamically unstable\citep{Valtonen2008}. Thus we expect the orbits to evolve on rapid timescales
(with respect to the expected stellar lifetime), 
especially as the disk dissipates. A natural outcome of this dynamical instability is the formation 
of a more hierarchical system with a tighter (few AU) inner 
pair and wider (100s to 1,000s AU) tertiary\citep{kiseleva1994},
consistent with observed triple systems\citep{tokovinin2011}. 

Our results demonstrate that protostellar disks can experience gravitational instability at 
young ages, leading to the formation of hierarchical multiple systems. Such disks 
may also show spiral structure associated with the linear growth phase of gravitational instability.
The spiral structure in L1448 IRS3B is consistent with
simulated ALMA observations of unstable disks that assume a comparable ratio of disk mass 
to stellar mass\citep{dipierro2014}. Spiral 
structure has been detected previously in more-evolved disks from observations of 
near-infrared scattered light originating from the disk upper layers\citep{grady2013}. However,
those disks are not sufficiently massive to generate the spiral structure by gravitational instability,
nor is the spiral structure apparent in observations of the disk midplanes. All these cases are 
more consistent with spiral arms triggered by a low-mass stellar, brown dwarf, or planetary mass 
companion\citep{dong2016}. 

Although the frequency of fragmenting disks like IRS3B is still unknown,
we do not expect them to be intrinsically rare. Our conclusion is 
based on the typical multiplicity and separation distributions observed 
in the parent star forming region, Perseus. Close multiples such as the 
IRS3-a/b or IRS3B-a/c pairs are frequently found in protostar 
systems\citep{tobin2016,connelley2008,takakuwa2012}.
This discovery, together with the recent work supporting turbulent 
fragmentation at larger scales (1,000s of AU)\citep{pineda2015,lee2016}, provide direct
observational support for the two dominant theoretical models for binary/multiple star
formation. We predict that systems similar to IRS3B 
will be discovered in larger samples with facilities like ALMA, and their frequency
will help benchmark the relative contribution of disk fragmentation to the population 
of binary/multiple stars.

\clearpage

Acknowledgments:
J.J.T acknowledges support from the University of Oklahoma, the Homer L. Dodge
endowed chair, and grant 639.041.439 from the Netherlands
Organisation for Scientific Research (NWO). K.M.K. is supported in part 
by the National Science Foundation under Grant No. AST-1410174.
M.V.P. is supported by ERC consolidator grant 614264 and A-ERC grant 291141 CHEMPLAN.
D.S.-C. acknowledges support for this work was provided by 
the NSF through award SOSPA2-021 from the NRAO.
Z.-Y.L. is supported in part by NSF AST1313083 and NASA NNX14AB38G.
This paper makes use of the following ALMA data: ADS/JAO.ALMA\#2013.1.00031.S.
ALMA is a partnership of ESO (representing its member states), NSF (USA) and 
NINS (Japan), together with NRC (Canada), NSC and ASIAA (Taiwan), and 
KASI (Republic of Korea), in cooperation with the Republic of Chile. 
The Joint ALMA Observatory is operated by ESO, AUI/NRAO and NAOJ.
The National Radio Astronomy 
Observatory is a facility of the National Science Foundation 
operated under cooperative agreement by Associated Universities, Inc.
This research made use of APLpy, an open-source plotting package for Python 
hosted at http://aplpy.github.com. This research made use of Astropy, 
a community-developed core Python package for 
Astronomy (Astropy Collaboration, 2013) http://www.astropy.org.

Contributions:
J.J.T., M.M.D., L.W.L., D.S.-C., C.J.C., L.P., C.M., S.I.S.,and R.J.H. participated in data acquisition, and
J.J.T., K.M.K., and M.V.P. contributed in data analysis and reduction.
J.J.T and K.M.K led the writing of the manuscript, incorporating the feedback, suggestions, and discussions of
results from all authors.

Reprints and permissions information is available at www.nature.com/reprints

Competing financial interests:
The authors declare no competing financial interests.

Corresponding Author:
Correspondence and requests for materials should be addressed to John J. Tobin; jjtobin@ou.edu.

\begin{figure}[!ht]
\begin{center}
\includegraphics[scale=0.55]{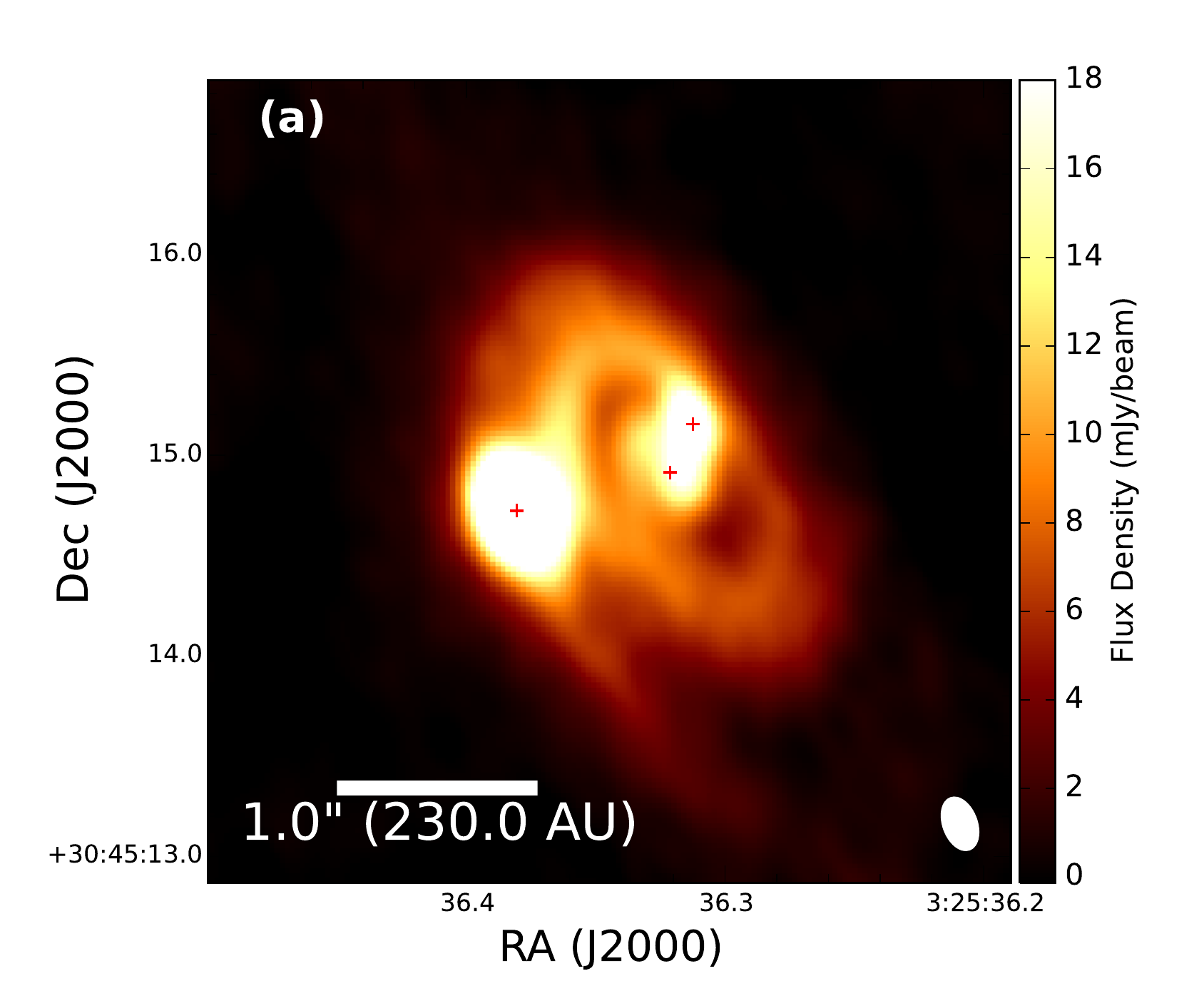}
\includegraphics[scale=0.55]{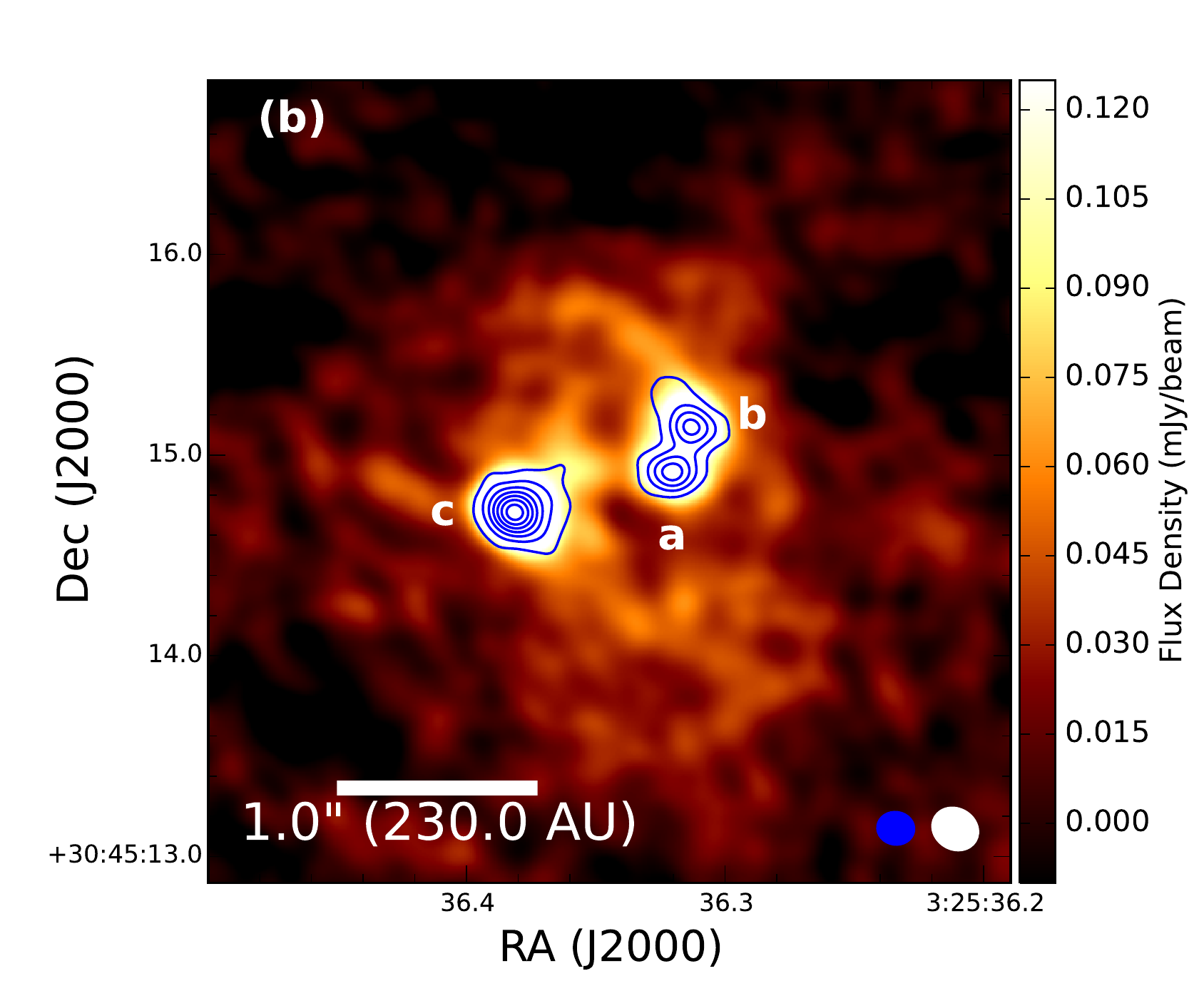}
\end{center}
\caption{ALMA and VLA images of the disk and triple protostar system L1448 IRS3B.
(a) ALMA 1.3~mm image of the extended disk, showing an evident bright source on the left (IRS3B-c)
 in the outer disk and another blended source on the right near the 
disk center (IRS3B-a and IRS3B-b). (b) VLA 8~mm image smoothed to a similar
resolution as the ALMA image, capturing some of the 
faint, extended  disk at longer wavelengths. The contours in panel (b) are from a 
higher-resolution VLA 8~mm image\citep{tobin2016} clearly 
showing the individual protostars with corresponding designations. 
All three protostars are embedded within apparent spiral arms that emerge from
IRS3B-a/IRS3B-b and extend to IRS3B-c in the outer disk. The positions of the three protostars 
identified from the VLA data are shown by red crosses in panel (a).
The contours start at and increase with 5$\sigma$, where $\sigma$ = 0.009 mJy beam$^{-1}$. The resolution of
each image is shown with an ellipse(s) drawn in the lower right corner, corresponding to 
0\farcs27$\times$0\farcs16 (62~AU~$\times$~37~AU) for the ALMA image in panel (a), 
0\farcs24$\times$0\farcs20 (55~AU~$\times$~46~AU) for the VLA image in panel (b), and
0\farcs18$\times$0\farcs16 (41~AU~$\times$~37~AU; blue ellipse) for the contour image in panel (b).
}
\label{diskcontinuum}
\end{figure}

\begin{figure}[!ht]
\begin{center}
\includegraphics[scale=0.5]{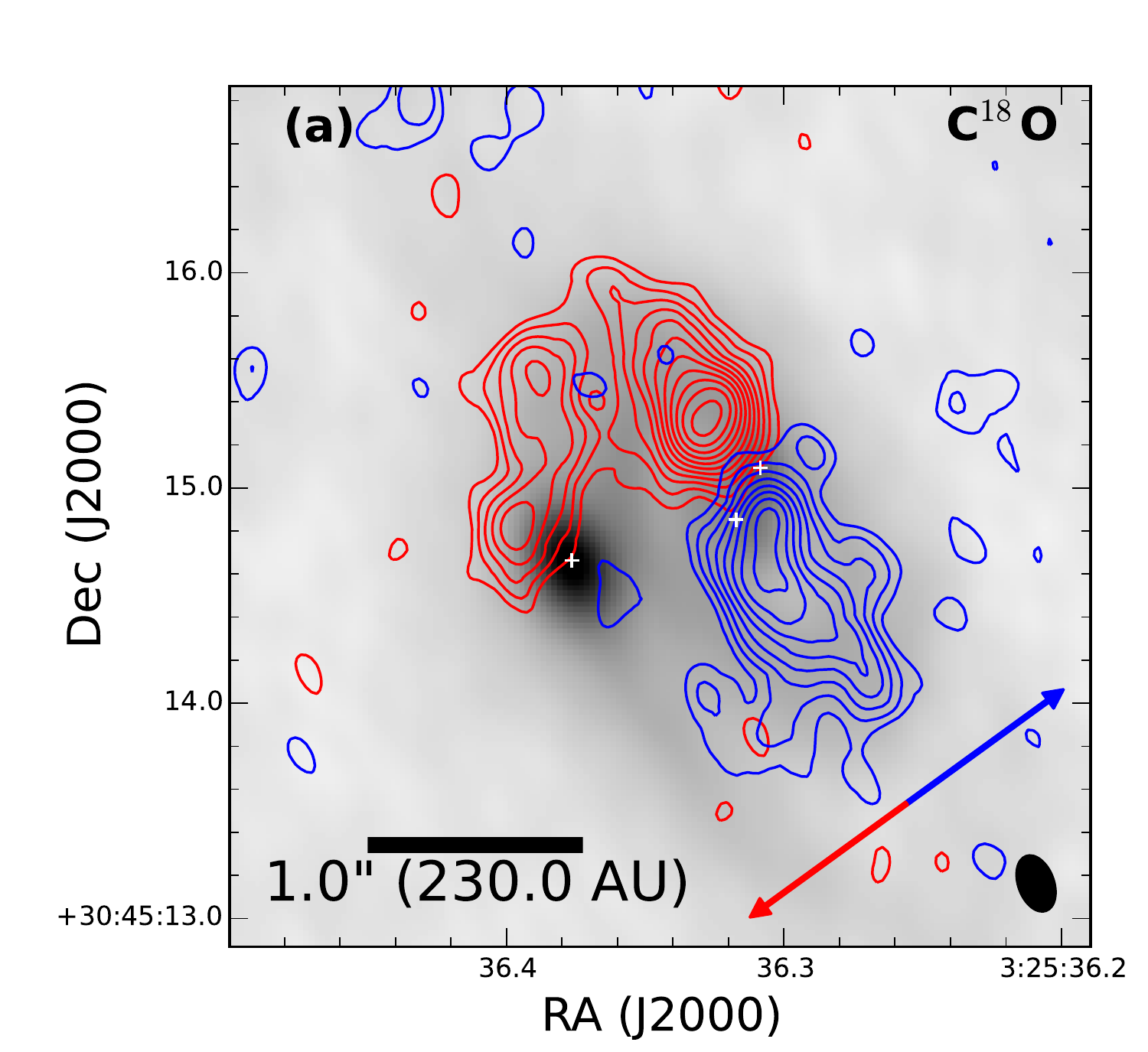}
\includegraphics[scale=0.5]{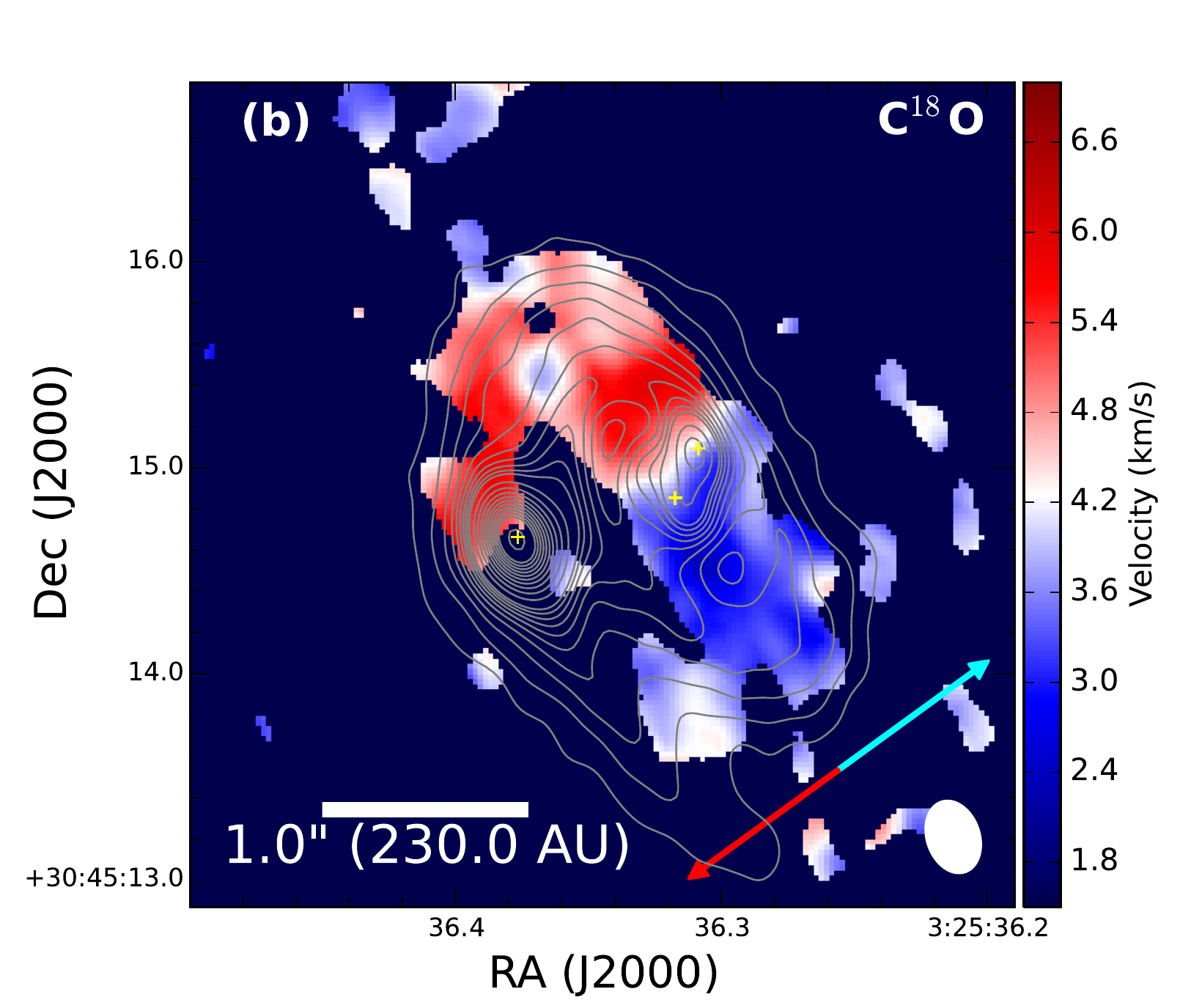}
\end{center}
\caption{Images of the C$^{18}$O emission and its corresponding velocity maps
from the disk around L1448 IRS3B showing a rotation signature.
(a) Red and blue-shifted C$^{18}$O ($J=2\rightarrow1$) emission are
overlaid on the ALMA 1.3 mm continuum image (grayscale) 
as red and blue contours. (b) Line-center velocity map 
of the C$^{18}$O emission with 1.3 mm continuum contours overlaid in gray.
The C$^{18}$O traces higher-velocity emission near IRS3B-a and IRS3B-b, consistent
with the system center of mass being near these protostars.
Lower velocity gas is found to be associated with
the outer spiral arm detected in dust emission.
The molecular line emission does not fully trace the disk 
due to spatial filtering of emission with velocities close to that 
of the system ($\sim$ 4.5~\kms). The source positions are marked with white or yellow crosses. 
The outflow direction\citep{lee2015} is denoted by the blue and red arrows.
The angular resolution of these data is given by the ellipse in the 
lower right corners, 0\farcs36$\times$0\farcs25 (83~AU~$\times$~58~AU).
The contours in panel (a) start at 4$\sigma$ and increase in 1$\sigma$ intervals. The
C$^{18}$O emission was integrated over 1.25 - 4.0~\kms\ and 5.5 - 7.0~\kms\ 
for the blue and red-shifted maps, respectively. The noise levels for 
C$^{18}$O are $\sigma_{Blue}$=2.25~K~\kms\ and $\sigma_{Red}$=1.65~K~\kms. 
The continuum (gray) contours in panel (b) start at and increase by 10$\sigma$, 
at 100$\sigma$ the levels increase in steps of 30$\sigma$, and at 400$\sigma$ 
the levels increase by 100$\sigma$ steps; $\sigma$=0.14 mJy~beam$^{-1}$.
} 
\label{diskkinematics}
\end{figure}

\begin{figure}[!ht]
\begin{center}
\includegraphics[scale=1.2]{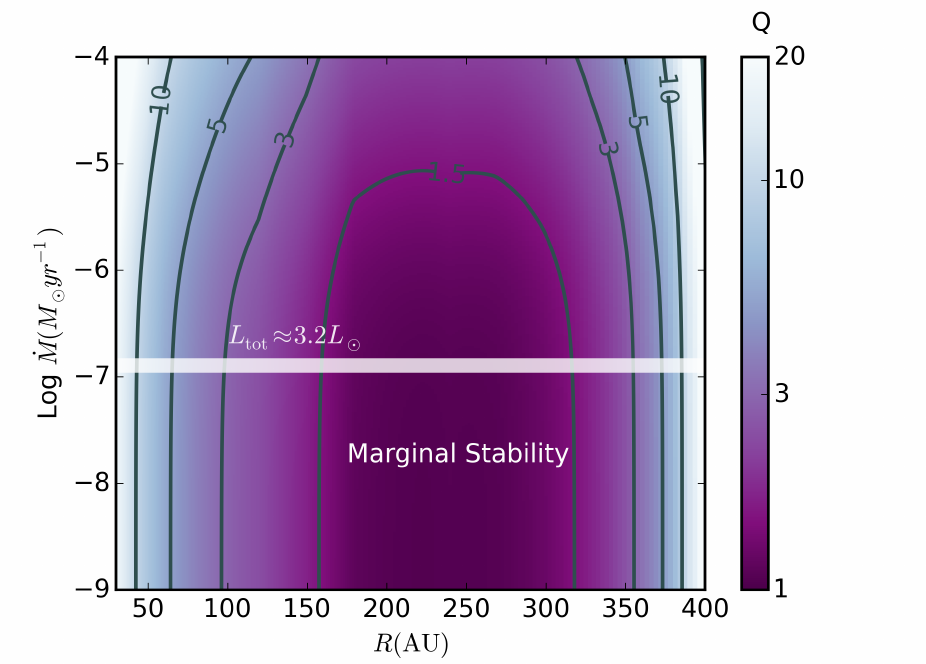}
\end{center}
\caption{
Plot of Toomre Q (instability criterion; Q $\approx$ 1 or less being unstable), versus
mass accretion rate and radius. The y-axis is the mass accretion rate in the 
disk and the gray contours are Q at a given radius and accretion rate. The luminosity and estimated
1.0~$M_{\sun}$ mass of the inner pair indicates a  mass accretion rate of $\sim$10$^{-7}$ M$_{\odot}$ 
yr$^{-1}$ (thick horizontal line).  Q approaches $\sim$1 at radii between 150--300~AU 
indicating that the outer disk is marginally unstable. At radii $>$350~AU
the surface density of the disk declines, such that it becomes
stable again. However, the outer radius of $\sim$400~AU, (800~AU (3\farcs5) diameter), is comparable to
the spatial scale at which our observations no longer recover all emission.
} 
\label{GImodels}
\end{figure}

\clearpage

\section{Methods}

\section{1. Observations}

L1448 IRS3B was observed with the Atacama Large Millimeter/submillimeter Array (ALMA) during
 Cycle 2 on 27 September 2015 with 33 antennas sampling baselines between 32 - 2000 meters. 
The observations were executed within a 1.8 hour block
and the total time spent on L1448 IRS3B was $\sim$2.9 minutes. 
The precipitable water vapor was $\sim$0.7 mm throughout the observing session.
The phase calibrator was J0319+4130 (3C84), the bandpass calibrator
was J0237+2848, and the amplitude and absolute flux calibrator was the
monitored quasar J0238+166. 
The correlator was configured to observe a 2 GHz continuum band centered at 232.5 GHz, and with 60 MHz bands centered
on the following molecular transitions: $^{12}$CO ($J=2\rightarrow1$), 
$^{13}$CO ($J=2\rightarrow1$), C$^{18}$O ($J=2\rightarrow1$), SO ($J_N = 6_5\rightarrow5_4$), 
and H$_2$CO ($J=3_{03}\rightarrow2_{02}$). The flux calibration accuracy is expected to be better than 10\%.

The raw data were manually reduced
by the North American ALMA Regional Center staff using CASA\citep{mcmullin2007} version 4.5.0.  We performed
self-calibration on the continuum data to increase the signal to noise ratio by correcting for short timescale
phase and amplitude fluctuations. The phase and amplitude solutions
from the continuum self-calibration were also applied to the 
spectral line bands. The 
data were imaged using the \textit{clean} task within CASA 4.5.0; the ALMA images shown 
in Figure 1 were generated using Briggs weighting with a robust parameter of 0.5.
Within the \textit{clean} task, we only include data at $uv$-distances $>$ 50 k$\lambda$ to 
mitigate striping in the images from large-scale emission detected on the shortest baseline
that could not be properly imaged. The molecular line images
shown in Figure 2 (main text) and Extended Data Figures 1 and 2 
were imaged with natural weighting, tapering at 500 k$\lambda$, and with data having
$uv$-distances $>$ 50 k$\lambda$. Tapering reduces the weight of longer 
baseline data in the deconvolution to facilitate the detection of larger, lower surface brightness
 structures. The beam size of the continuum image is
0\farcs27$\times$0\farcs16 (62~AU~$\times$37~AU) and the beam size of 
the molecular line data is 0\farcs36$\times$0\farcs25 (83~AU~$\times$~58~AU). 
The resultant noise in the 1.3~mm continuum 
was 0.14~mJy/beam and 15~mJy/beam in 0.25~\kms\ 
channels for the spectral line observations.

We show the $^{12}$CO ($J=2\rightarrow1$) and 
H$_2$CO ($J=3_{03}\rightarrow2_{02}$) integrated intensity maps
in Extended Data Figure 1. The $^{12}$CO map clearly shows an outflow(s) originating from the region
of the three protostars.
The red-shifted emission shows a clear, wide outflow cavity, with another red-shifted feature within it. We 
suggest that this secondary red-shifted feature within the outflow cavity is a jet-like outflow
from IRS3B-c and that the
arc-like shape is due to the orbital motion of the source. The blue-shifted outflow is much
more diffuse and not as well-recovered in our data, but there appears to be emission associated
with all three protostars. Nonetheless, it is clear that the outflow map toward these three sources
is quite complex and made even more complex by the two more widely separated systems (IRS3A and IRS3C)\citep{looney2000}
that also drive outflows that are extending across this map. The larger-scale 
outflow has been examined more thoroughly at lower resolution\citep{lee2015}.

We also detect faint H$_2$CO emission from the vicinity of L1448 IRS3B, shown in Extended Data Figure 1. The emission
has low intensity, but also shows a velocity gradient in the same direction as $^{13}$CO and C$^{18}$O. 
SO ($J_N = 6_5\rightarrow5_4$) emission was not detected toward L1448 IRS3B. H$_2$CO is expected to
trace the inner envelope and disk around the protostars\citep{sakai2014b} and it is also
sometimes present in outflows \citep{oya2014}. SO has been found to trace
both outflow shocks\citep{wakelam2005} and possibly a centrifugal barrier, highlighting a transition
between the disk and infalling envelope of protostar\citep{sakai2014a,oya2016}.

The $^{13}$CO ($J=2\rightarrow1$) emission traces kinematics similar to that of the higher-velocity C$^{18}$O, with the
blue and red-shifted emission concentrated around IRS3B-a and IRS3B-b. The integrated intensity maps of the
red and blue-shifted emission are shown overlaid on the ALMA 1.3~mm continuum image in Extended Data Figure 2, along
with the velocity map derived from the $^{13}$CO emission. The $^{13}$CO does not extend toward 
IRS3B-c because of spatial filtering; $^{13}$CO is more severely affected by spatial filtering because it is 
more abundant and has higher opacity than C$^{18}$O, leading to a wider velocity range that is filtered-out. There is
some outflow emission detected in the $^{13}$CO, but the apparent rotation of the inner disk(s)
 around IRS3B-a and IRS3B-b dominates the velocity field for the detected $^{13}$CO. Previous studies have
also found that $^{13}$CO can be a good tracer of the disk kinematics with sufficient spatial resolution to
rule-out outflow contamination\citep{takakuwa2012,tobin2012,harsono2014}.

\section{2. Subtracting the Tertiary, IRS3B-c}

To characterize the circumtriple disk, it is necessary to remove the 
bright protostar IRS3B-c located in the outer arm of the disk. 
IRS3B-c has the largest peak intensity in the system and its presence 
potentially masks some of the underlying disk structure, in addition to skewing the derivation
of system geometry (e.g., inclination, position angle). Furthermore, IRS3B-c appears internally
heated and is likely optically thick and the process of removing it from the data enables us to
characterize its properties, independently of the disk that it is embedded within.


To remove IRS3B-c as cleanly as possible, we fit a two-component Gaussian with
a constant zero-level offset. The two-component Gaussian is comprised of a point-source component
superposed with a more extended Gaussian, and the 
zero-level offset is used to preserve emission from the disk in the vicinity of
IRS3B-c source. The fitting region is also restricted to a 
0\farcs7 $\times$ 0\farcs6 ellipse around IRS3B-c, ensuring 
that the Gaussian fit to IRS3B-c is not
strongly affected by the emission in the surrounding, extended disk.

We used the \textit{imfit} task in CASA 4.5.0 to fit the Gaussian components. We then used these components to 
construct a model image of IRS3B-c within CASA. Next, we used the \textit{setjy} task to generate
visibility data from the model image via a Fourier transform, and 
then filled the model column of the CASA measurement set
with the visibility data. Finally, we use the \textit{uvsub} task to 
subtract the model column from the data column, producing
a dataset with IRS3B-c subtracted. Images with IRS3B-c removed
were then generated with \textit{clean}, and we show the resultant images
in Extended Data Figure 3.

With IRS3B-c removed, we fit the major and minor axes of the disk to estimate the 
inclination with respect to the line of sight and position angle. To do this, we 
first generated an image from the dataset with IRS3B-c subtracted using the CASA \textit{clean} 
task, but taper (e.g., smooth, see Section 1) the visibility data to have a resolution of
 0\farcs43$\times$0\farcs36; this image smooths over the spiral structure
such that a single Gaussian can better fit the disk emission from the entire source. 
We use the \textit{imfit} task in CASA to fit a single 
Gaussian, with the center position fixed at the location of IRS3B-a, which appears to
be the most centrally-located source
in the system.

The deconvolved major and minor axes of the resulting 
Gaussian were 1\farcs78 and 1\farcs25 (409~AU~$\times$~288 AU). Assuming symmetry, this 
corresponds to an inclination angle of 
45.4\degr$^{+10.4}_{-12.8}$ ($arccos(\theta_b/\theta_a$)) and the 
position angle is 29.5\degr\ east of north. The uncertainty in the
inclination results from the disk being asymmetric, as shown by the 
Figure 1 in the main text and Extended Data Figure 3 (see Section 4).
We estimate that there could be a systematic uncertainty of 20\% in the minor axis of the
disk because of the disk asymmetry.


\section{3. Disk Mass}
The dust emission from L1448 IRS3B can be used to estimate the mass of the surrounding
disk and IRS3B-c itself. Assuming that the dust around L1448 IRS3B is
optically thin, isothermal, and that the gas and dust are well-mixed, 
we can calculate the mass with the equation
\begin{equation}
\label{eq:dustm}
M_{dust} = \frac{D^2 F_{\lambda} }{ \kappa_{\lambda}B_{\lambda}(T_{dust}) }.
\end{equation}
F$_{\lambda}$ is the integrated flux density at 1.3~mm, B$_{\lambda}$ is the Planck function,
and $D$ = 230 pc\citep{hirota2011} is the distance to the Perseus molecular cloud. We adopt
$T_{dust}$~=~30~K as a typical dust temperature at a radius of 100 AU, derived from radiative
transfer models\citep{jorgensen2009,tobin2013}. The disk does have a temperature gradient, but 30 K
is a reasonable average value for the outer disk where most of the mass resides.
The remaining term $\kappa_{\lambda}$ is the dust opacity at the
observed wavelength, which is adopted from  dust opacity models. While there are a multitude of possible
models to consider\citep{testi2014}; we adopt two limiting cases for simplicity.
The first value is $\kappa_{1.3mm}$~=~0.899 cm$^2$~g$^{-1}$[\citep{ossenkopf1994}], 
appropriate for dense cores in molecular clouds like L1448 IRS3B. The second value we
adopt is $\kappa_{1.3mm}$~=~2.3 cm$^2$~g$^{-1}$, which is typically used for proto-planetary disks 
\citep{andrews2010,ansdell2016}. The second value is derived from a parametrization of the dust opacity
to have a value of 10.0 cm$^2$~g$^{-1}$ at $\lambda$~=~0.3~mm and 
$\kappa_{\lambda}$~=~10.0~$\times$~($\lambda$/0.3~mm)$^{-\beta}$, 
where $\beta$~=~1[\citep{beckwith1990, andrews2010, ansdell2016}]. 
We then adopt a standard dust to gas mass ratio of 1:100[\citep{bohlin1978}] and assume 
that this ratio is constant throughout the disk.
We note that the dust to gas ratio may be changing in more-evolved 
proto-planetary disks\citep{bergin2013,williams2014,ansdell2016},
but we assume the constant, canonical value for this much younger system.
Additionally, young protostellar disks may not be optically thin at 1.3~mm, and thus our mass estimates
using Equation \ref{eq:dustm} are likely lower limits. Note that a 
higher disk mass would only strengthen our conclusion that 
the disk is unstable. We show below that analytic models of the disk suggest that the inner parts 
are marginally optically thick at somewhat shorter wavelengths.

Keeping the assumptions and caveats from the previous paragraph in mind, we can 
now calculate the masses of the disk surrounding the three protostars.
We calculate a total mass of 0.39 M$_{\sun}$ for the disk surrounding L1448 IRS 3B, 
assuming $\kappa_{1.3mm}$~=~0.899 cm$^2$~g$^{-1}$. 
Of the total, 0.085 M$_{\sun}$ is contained
within the concentrated dust around IRS3B-c; note that we adopt 
$T_{dust}$~=~40~K for IRS3B-c since the region directly
around it should be warmer and the peak brightness temperature 
(T$_B$) in the data is $\sim$40~K toward this source.
This high peak T$_B$ is indicative of increased opacity in IRS3B-c
because T$_B$ = T when the emission source is optically thick. 
Thus, the mass estimate of 
0.085~M$_{\sun}$ is also likely to be a lower limit because of optical depth effects.
If we instead adopt $\kappa_{1.3mm}$~=~2.3 cm$^2$~g$^{-1}$, assuming that the dust 
has already grown substantially from ISM size as in older proto-planetary disks, we find a total mass 
of 0.15 M$_{\sun}$ (disk plus IRS3B-c) and 0.03 M$_{\sun}$ for IRS3B-c alone. 

To conclude, the higher mass estimates using $\kappa_{1.3mm}$~=~0.899 cm$^2$~g$^{-1}$ are
favored because that opacity model is meant to reflect the dust typically found in dense molecular 
clouds. Since L1448 IRS3B is still in a very early stage of the star formation process,
the bulk of its dust should reflect the molecular cloud content rather than the dust in 
proto-planetary disks that has been able to undergo significant dust evolution\citep{perez2012,testi2014}.

\section{4. Disk Surface Density}

The asymmetry and complexity of the dust continuum around L1448 IRS3B prohibit us from directly fitting
symmetric disk models to the data to determine the underlying density structure. 
 However, given that the disk is well-resolved, we can empirically determine
the azimuthally averaged surface density profile as a function of radius from the data. We first deprojected the 
visibility data using the inclination of 45.4\degr\ and position angle of 29.5\degr, determined from the Gaussian 
fitting in Section 2. The uv-data are deprojected using standard methods\citep{persson2016} that adjust
the $u$ and $v$ coordinates based on the geometric projection.
We then generate deprojected images using the CASA \textit{clean} task in the same manner
as we made images for the non-deprojected data (see Section 1). The deprojected image is shown 
in Extended Data Figure 3.

With the deprojected image, we measure the flux density in a series of 
circular annuli, centered on IRS3B-a; each annulus
has a width of half the beam-size. Afterwards, we divide the flux density 
in each annulus by its surface area and 
converted the surface brightness to a surface density using the methodology described in Section 3,
assuming a radial temperature profile that is described in Section 6. The surface density ($\Sigma$)
inferred from the observed intensity profile, and the assumed temperature profile 
are shown in Extended Data Figure 4.

\section{5. Protostar Masses}

It is essential to know the masses of the central protostars (IRS3B-a and IRS3B-b) to quantify the 
stability of the disk. The angular resolution of the observations is not sufficient to
determine the masses of each component of the central pair, and our measurement will be limited
to the combined mass of both components. Figure 2 of the main text shows 
the integrated intensity and velocity maps of the
C$^{18}$O ($J=2\rightarrow1$) emission and the $^{13}$CO ($J=2\rightarrow1$) 
is shown in Extended Data Figure 2. The maps
show clear evidence of rotation, centered on IRS3B-a and IRS3B-b, based on
the spatial separation of the blue and red-shifted components. Furthermore, the C$^{18}$O emission
is also tracing some of the structure observed in dust emission in the outer disk.

To examine the molecular line kinematics, we extract position-velocity (PV) diagrams along the major axis of
the disk, centered on IRS3B-a and IRS3B-b. The PV diagram for C$^{18}$O is shown in Extended Data Figure 5.
The signal-to-noise ratio of the data is not high enough to employ the techniques that have been used to
quantitatively measure protostellar masses in other 
works\citep{tobin2012,ohashi2014,seifried2016,oya2014,oya2015,oya2016}.
Nevertheless, we can estimate the protostar mass by drawing a Keplerian curve on the PV
diagram, accounting for the 45.4\degr\ system inclination. We find that the
combined mass for both IRS3B-a and IRS3B-b
is likely $\sim$1.0~M$_{\sun}$. We also compare
this estimate to a PV diagram of a rotating thin disk model\citep{maret2015}
with the same central mass and inclination, shown in Extended Data Figure 4. The thin disk model 
occupies a very similar domain in PV space as the data. Notice that the line
we draw in the PV diagrams does not go through the middle of the data, 
but instead traces the edge of the highest velocity emission for a given 
position. This is because observations of a Keplerian disk will 
measure multiple velocities superposed at a given 
position and the Keplerian velocity at a given radius only 
corresponds to the highest velocities\citep{seifried2016}. Thus, the 
total protostellar mass cannot be substantially larger than 1.0 M$_{\sun}$ 
without being inconsistent with the data.

\section{6. Stability Analysis}

The simplest estimate for when disks are subject to gravitational instability is provided by the
Toomre Parameter\citep{toomre1964}:
\begin{equation}\label{eq:Toomre}
Q = \frac{c_{s}\kappa}{\pi G \Sigma},
\end{equation}
where $c_s$ is the sound speed within the disk, 
$\kappa$ is the epicyclic 
frequency (which is $\Omega$, the angular rotation frequency, for Keplerian disks), $\Sigma$ is 
the disk surface density, and $G$ is the gravitational 
constant. The quantities $c_s$, $\Omega$, and $\Sigma$ are all 
functions of radius within a disk. If $Q$ reaches order unity, 
the disk becomes susceptible to the growth of spiral
density waves. Disk fragmentation due to gravitational instability is 
thought to be the non-linear outcome of this hydrodynamic
instability\citep{kratter2016}. The evolution of disks subject to 
the instability, and in particular their ability to create long-lived, bound objects,
 is a function of many parameters, including infall
from the environment and radiative heating and cooling. The 
importance of the latter effect can be reasonably well encapsulated by the disk cooling time:
\begin{equation}
\label{eq:tcooldisk}
t_{\rm cool} = \beta \Omega^{-1} = \frac{4}{9\gamma(\gamma-1)}\frac{\Sigma c_{\rm s}^2}{\sigma T^4}f(\tau).
\end{equation}
where
\begin{equation}\label{eq-ftau}
f(\tau)=\tau +\frac{1}{\tau},
\end{equation}
where the optical depth $\tau = \kappa\Sigma/2$ is calculated using the Rosseland mean 
opacity. $f(\tau)$ reasonably captures how the accretion energy diffuses 
from the midplane in the optically thick and thin regimes\citep{rafi2005}.

For fragments to survive, the gas must be able to contract by 
radiating away heat generated by shocks on roughly the orbital 
timescale \citep{gammie2001}. $\beta$ is often used in the 
literature to parametrize cooling via a single dimensionless number to probe disk behavior in 
different regimes. Although there is some debate about the critical value of $\beta$ 
below which fragmentation successfully produces bound objects\citep{meru2011,paardekooper2012,lodato2011}, 
fragment formation is robust for $\beta < 10$. 

To assess whether IRS3B is a good candidate for disk fragmentation, 
we generate a set of one-dimensional analytic disk models. Specifically,
we calculate the Toomre parameter and cooling time as a function of disk radius. 
We use the surface density 
profile derived for the disk after the removal of IRS3B-c (see Section 4), and 
treat IRS3B-a and IRS3B-b as equal mass 0.5~\Msun\ protostars (see Section 5). 
We use the youngest age available from 
the Baraffe et al. models\cite{baraffe2015} (1 Myr) to calculate the stellar radii and intrinsic 
luminosity. To estimate the disk temperatures, a two 
component temperature model is used\citep{kratter2008,kratter2010}. The disk midplane temperature 
is found by balancing heating and cooling terms:
\begin{equation}\label{eq-disktemp}
\sigma T^4 = \frac{3}{8}f(\tau) F_{\rm acc} + F_*. 
\end{equation}
Here $F_*$ represents the energy flux associated with irradiation from the protostars,
\begin{equation}\label{eq-mlirrad}
F_{*}  = \sigma T_{*}^4= f \frac{L_*}{4\pi r^2}
\end{equation}
where $f\approx 0.1$ is based on ray tracing calculations that show that much of the stellar 
flux impacts the infalling envelope\citep{tsc1984} and is reprocessed back down onto the disk 
in embedded sources\citep{matzner2005}. $L_*$ includes both the model stellar luminosity and 
the accretion luminosity, which are comparable to each other. $F_{\rm acc}$ is the energy flux due to 
dissipation of accretion energy within the disk:
\begin{equation}\label{eq-facc}
F_{\rm acc} = \frac{3}{8\pi}{\dot{M}\Omega^2}
\end{equation}

We consider two models for disk opacity: a temperature independent opacity of 
$\kappa$~= 0.24~cm$^2$~g$^{-1}$[\cite{pollack1985}], and an interpolated 
model, where roughly $\kappa$~$\propto$~T$^2$[\citep{semenov2003}]. The 
former model is more consistent with grain growth beyond mm sizes, while 
the latter should capture more ISM-like grains. For the disk models 
presented here the opacity has negligible effect on the disk temperature 
because it is dominated by stellar irradiation, which sets the disk surface 
temperature, and therefore is unaffected by optical depth. 
This is typical of unstable protostellar disks\citep{kratter2016}. When the midplane 
heating is relatively weak, the disk becomes nearly vertically isothermal \citep{chiang1997}. 
Note that the numerical values for these two models are distinct from those used to derive the disk mass 
in Section 2 because calculation of equilibrium temperatures analytically requires a mean opacity, rather than the 1.3~mm opacity.

We first evaluate the disk properties using temperature independent opacities, 
and explore the impact of a range of disk accretion rates. 
Recall that the disk accretion rate contributes both to the local 
viscous heating through $F_{\rm acc}$ and the stellar irradiation 
through $F_*$. Once we have calculated the disk temperature, we can 
then compute the Toomre parameter as a function of radius.
We also calculate the disk cooling time to demonstrate that if the disk is unstable, 
fragments that form could cool quickly and likely remain bound\citep{gammie2001,rice2005}. 
We find that the cooling time is well below the critical 
threshold of $\sim10\Omega^{-1}$ throughout the entire disk.
The discrepancy between our calculated $\beta$ values and those 
quoted in the literature derives mostly from the difference in 
radius between this disk and standard models with $r_d<100$AU.
Reaching $Q=1$ at larger radii requires lower values of $\Sigma$; 
$t_{\rm cool}$ scales roughly as $\Sigma^2$ in the optically thick 
regime relevant for such disks. In addition, the dimensionless cooling time
scales linearly with $\Omega$, which also declines at large radii.
In Figure 3 of the main text, we show contours of Q as a function of disk 
radius, and model accretion rate. The disk is marginally unstable between $150-320$~AU.

While we do not have a direct constraint on $\dot{M}$, the observed total luminosity 
provides a benchmark: In a typical viscous accretion disk, the total luminosity is:
\begin{equation}
L_{\rm tot} = L_{\star} + \frac{G M \dot{M}}{R_{\star}}
\end{equation}
where the second term represents the total accretion luminosity, half of which is 
liberated within the accretion disk, and half at the stellar surface. Thus we can 
derive the expected accretion rate assuming that
the bolometric luminosity of 3.2~L$_{\sun}$\citep{sadavoy2014} is
equivalent to the the internal luminosity. This implies an accretion 
rate of $\sim$10$^{-7}$~\Msun~yr$^{-1}$, which is consistent with star formation models\citep{shu1977} 
and observations of young stars\citep{enoch2009,offner2011}. This band is highlighted 
in Figure 3 of the main text. We account for the contributions from both protostars in the 
irradiation models, assuming that the accretion luminosity is divided equally between the two. 
Note that the internal luminosity is not precisely equivalent to the bolometric luminosity;
the bolometric luminosity is calculated from integrating the observed spectral energy distribution
from the near-infrared to the millimeter. To relate the bolometric luminosity 
to the internal luminosity, a correction must be applied to account for the 
geometric projection of the disk and outflow cavities. For an inclination of 45.4\degr\ the 
correction factor is $\sim$0.9[\citep{whitney2003a}]. Furthermore,
external irradiation can add an additional 0.1 - 0.5 L$_{\sun}$ to the bolometric luminosity 
(depending on the local environment); this amount should be subtracted from 
the bolometric luminosity to 
relate it to the internal luminosity. However, we do not apply these small corrections 
given that the uncertainty in the bolometric luminosity is $\sim$10\% or greater
depending on wavelength coverage\citep{enoch2009}, and these corrections would only lower the
internal luminosity (and by inference the accretion rate), causing Q to decrease and imply
a greater degree of instability.

We next calculate a disk model using a self-consistently derived temperature profile 
based on the temperature dependent opacities. For simplicity we only consider the benchmark 
accretion rate $\dot{M}$~$\approx$10$^{-7}$~\Msun~yr$^{-1}$. We create a grid of allowed temperatures 
and identify the radial temperature profile that is consistent with the radiative equilibrium 
model of Equation \ref{eq-disktemp}. In Extended Data Figure 4, we show the derived surface 
density, temperature, $Q$ and $\beta$ profiles. Once again we find that the disk is marginally
 unstable at radii in the vicinity of the recently formed tertiary, IRS3B-c. 

We note that although the surface density profile was derived using the optically 
thin approximation, our modeling indicates that the inner most regions may be marginally 
optically thick ($\tau$~$\sim$~2-5). Thus, we are likely underestimating the total disk
 mass somewhat. This would only corroborate our conclusion that the disk has most 
likely experienced a recent episode of gravitational instability and fragmentation. 

Finally, we can obtain a rough estimate of the mass scale of a fragment born in such a 
disk. The most unstable wavelength (i.e., physical scale) on which perturbations 
grow due to GI is $\lambda = 2 \pi H$, where H is the disk scale height (see Equation 1 of the main text). 
The typical initial fragment mass is, to order of magnitude, $M_{\rm frag} \approx \epsilon \Sigma \lambda^2$, 
where $\Sigma$ is the surface density of the disk. Various studies have debated the 
appropriate order unity coefficient $\epsilon$ to properly 
account for contraction and spiral arm overdensities\citep{boley2010}. For $\epsilon =0.5$, 
the fragment mass ranges from $0.005-0.015\Msun$ between $150-320$~AU for our disk model with an 
accretion rate of $10^{-7}$~\Msun~yr$^{-1}$.

Data Availability Statement: The ALMA data utilized to generate Figures 1 \& 2 in
the main text and Extended Date Figures 1, 2, 3, \& 5 are from ALMA project
2013.1.00031.S. These data will be available in the ALMA archive and from the corresponding
author upon request. The data from which Figure 3 of the main text was generated
and Extended Data Figure 4 are provided with the paper.

Code Availability: The data were reduced and processed using the Common Astronomy
Software Applications (CASA) version 4.5.0. The code for generating the model found 
in Extended Data Figure 5 is publicly available\citep{maret2015} http://dx.doi.org/10.5281/zenodo.13823. The theoretical calculations presented in Figure 3 of the main text
and in Extended Data Figure 4 are derived from solving algebraic equations 
and thus involve no significant code development.

\renewcommand{\figurename}{Extended Data Figure}
\renewcommand{\thefigure}{\arabic{figure}}

\setcounter{figure}{0}

\begin{figure}[!ht]
\begin{center}
\includegraphics[scale=0.5]{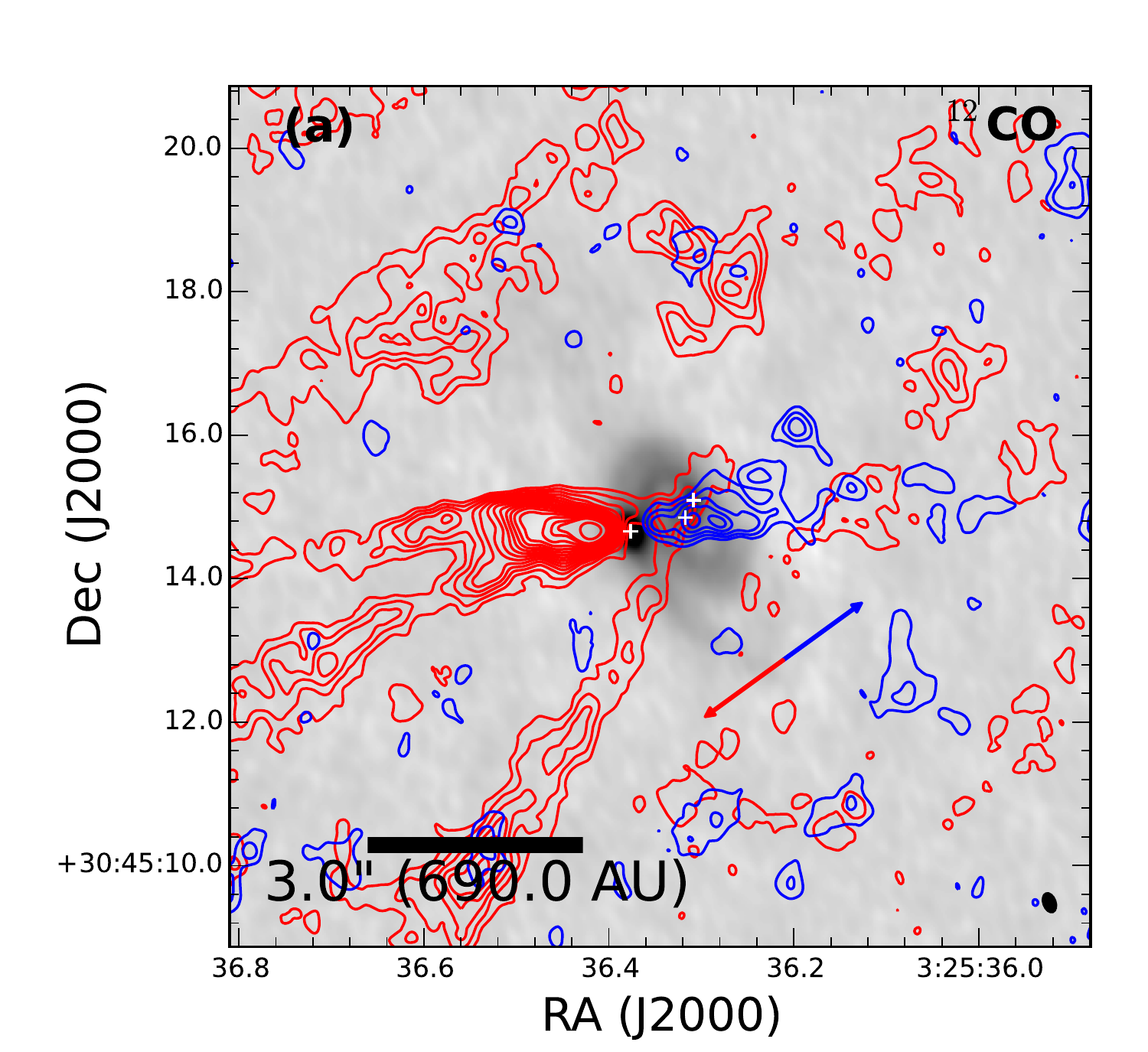}
\includegraphics[scale=0.5]{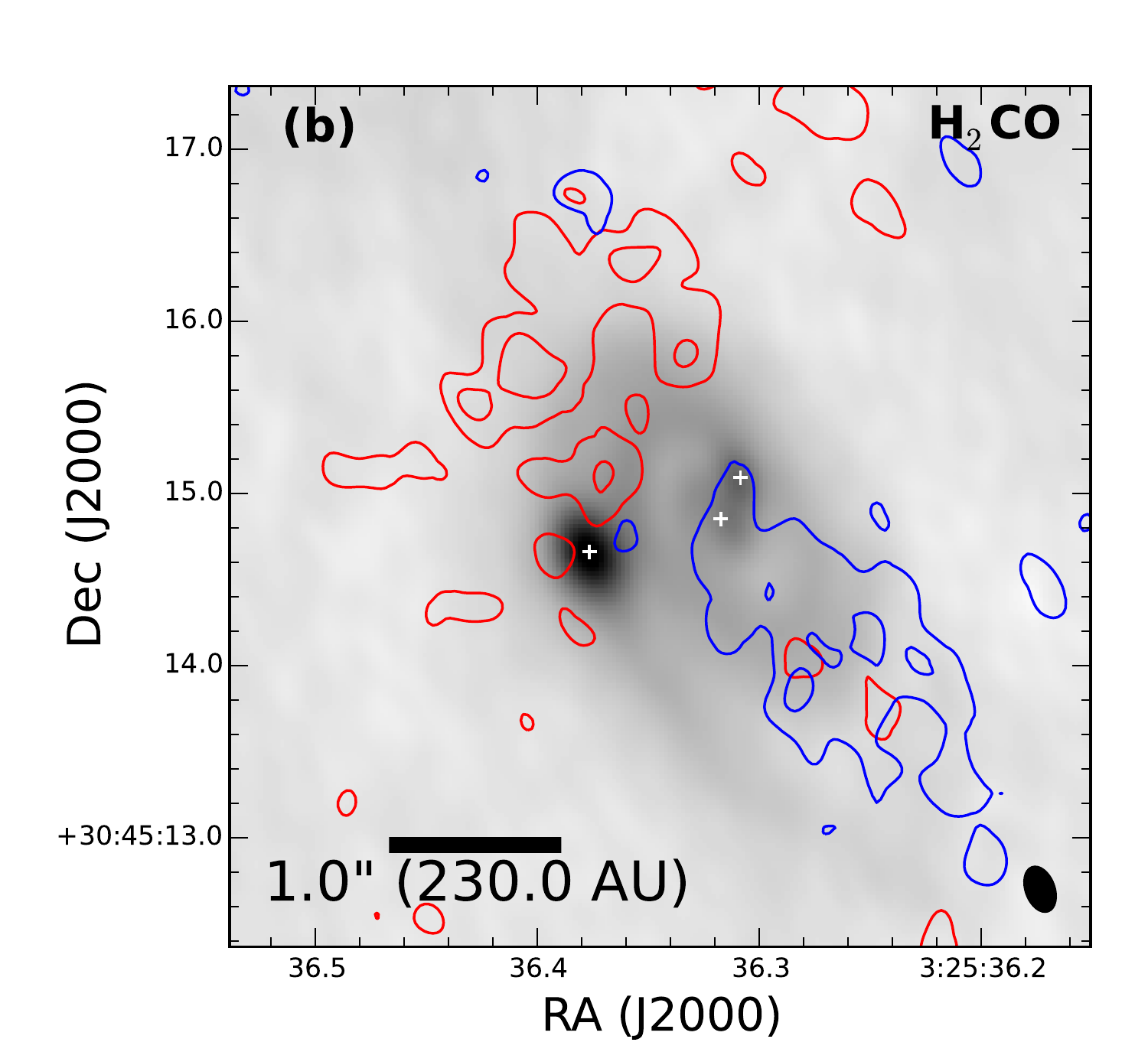}
\end{center}
\caption{Images of the $^{12}$CO and H$_2$CO emission in the vicinity of L1448 IRS3B. 
Panel (a) shows the $^{12}$CO red and blue-shifted contours overlaid on the 1.3~mm continuum
and the same is shown in panel (b) but for H$_2$CO. The $^{12}$CO emission in panel (a)
most clearly shows a red-shifted outflow from the 
three protostars. There is a wide cavity that is traced back to 
IRS3B-a/b and a more collimated outflow
is emitted from IRS3B-c, that is potentially generating the red-shifted arc that is
within the wide outflow cavity. The blue-shifted side of the outflow is more diffuse and not well-recovered
in our data, but appears associated with all three sources. The H$_2$CO emission in panel (b)
has low intensity and traces a rotation gradient in the inner envelope and disk surrounding the
sources. The source positions are marked with white crosses and the outflow direction is marked
with the blue and red arrows in the $^{12}$CO map.
 The contours in both plots start at 3$\sigma$ and increase in 2$\sigma$ intervals. The
$^{12}$CO emission was integrated over -5.5 - 1.5~\kms\ and 6.5 - 10.0~\kms\ for the blue
and red-shifted maps, respectively, and $\sigma_{Blue}$=6.88~K~\kms\ and $\sigma_{Red}=$5.02~K~\kms. The
H$_2$CO emission was integrated over 2.75 - 4.0~\kms\ and 5.25 - 6.25~\kms\ for the blue
and red-shifted maps, respectively, and $\sigma_{Blue}$=2.25~K~\kms\ and $\sigma_{Red}=$2.05~K~\kms. 
}
\end{figure}

\begin{figure}[!ht]
\begin{center}
\includegraphics[scale=0.5]{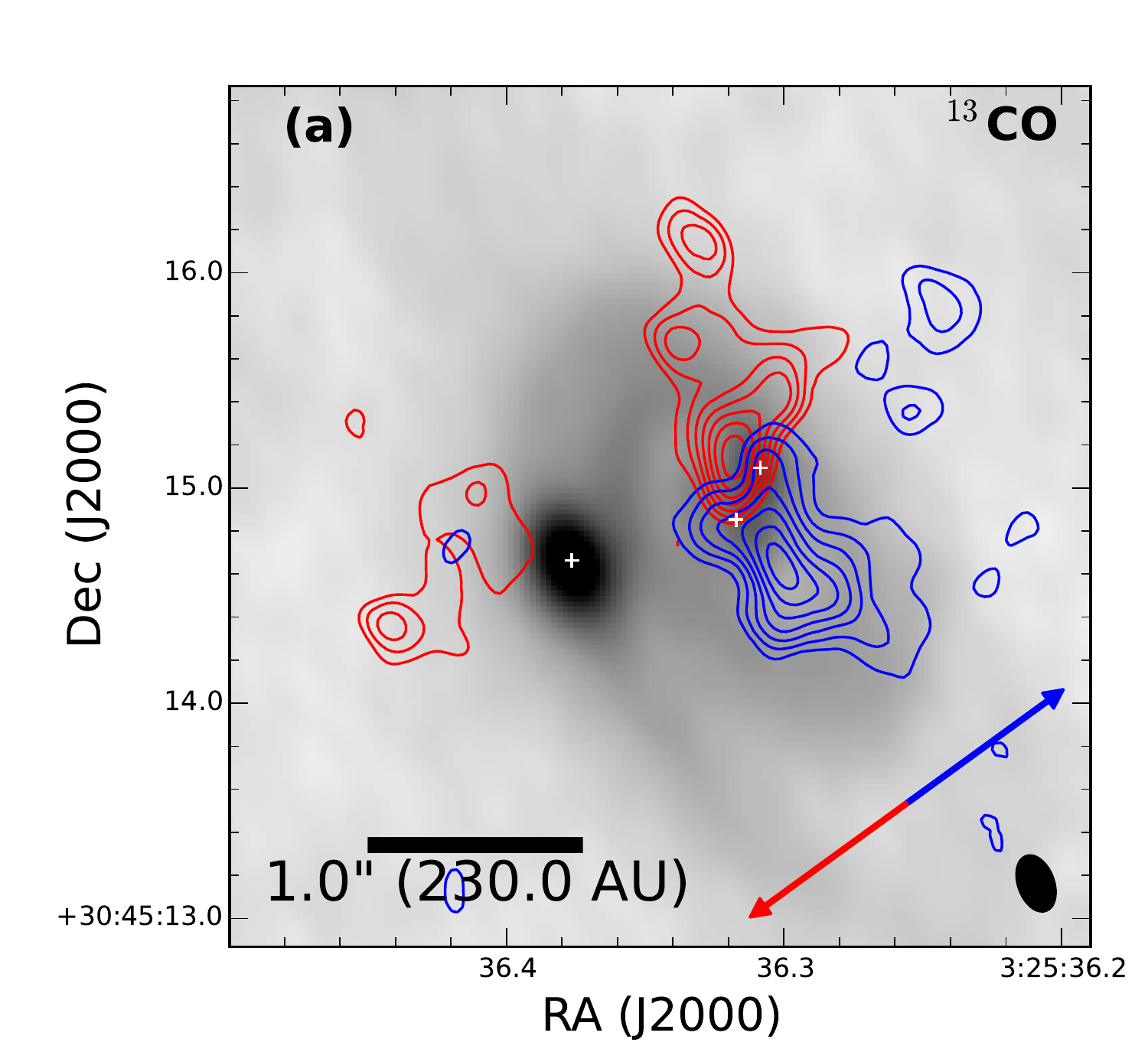}
\includegraphics[scale=0.5]{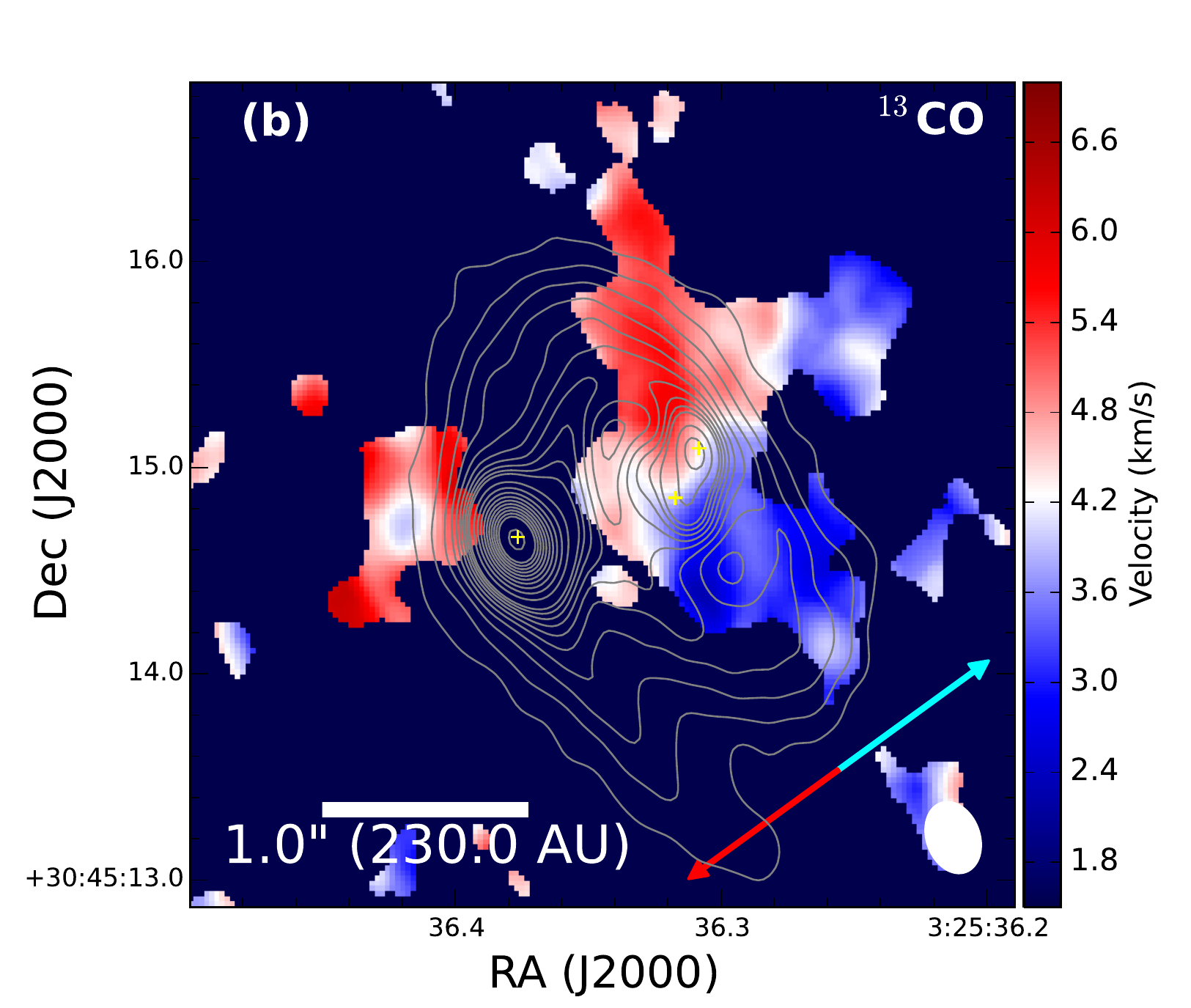}

\end{center}
\caption{Images of the $^{13}$CO emission and its corresponding velocity maps
from the disk around L1448 IRS3B, showing a rotation signature.
(a) Red and blue-shifted $^{13}$CO ($J=2\rightarrow1$) emission
overlaid on the ALMA 1.3 mm continuum image (grayscale) 
as red and blue contours. (b) Line-center velocity map 
of the $^{13}$CO emission with 1.3 mm continuum contours overlaid in gray.
The $^{13}$CO traces higher-velocity emission near IRS3B-a and IRS3B-b and
little of the extended disk due to spatial filtering.
The source positions are marked with white or yellow crosses. 
The outflow direction\citep{lee2015} is denoted by the blue and red arrows.
The angular resolution of these data are given by the ellipse in the 
lower right corners, 0\farcs36$\times$0\farcs25 (83~AU~$\times$~58~AU).
The contours in panel (a) start at 4$\sigma$ and increase in 1$\sigma$ intervals. The
$^{13}$CO emission was integrated over 1.25 - 4.0~\kms\ and 5.5 - 7.0~\kms\ 
for the blue and red-shifted maps, respectively. The noise levels for 
$^{13}$CO are $\sigma_{Blue}$=4.99~K~\kms and $\sigma_{Red}$=3.2~K~\kms. 
The continuum (gray) contours in panel (b) start at and increase by 10$\sigma$, 
at 100$\sigma$ the levels increase in steps of 30$\sigma$, and at 400$\sigma$ 
the levels increase by 100$\sigma$ steps; $\sigma$=0.14 mJy~beam$^{-1}$.
} 
\end{figure}

\begin{figure}[!ht]
\begin{center}
\includegraphics[scale=0.7]{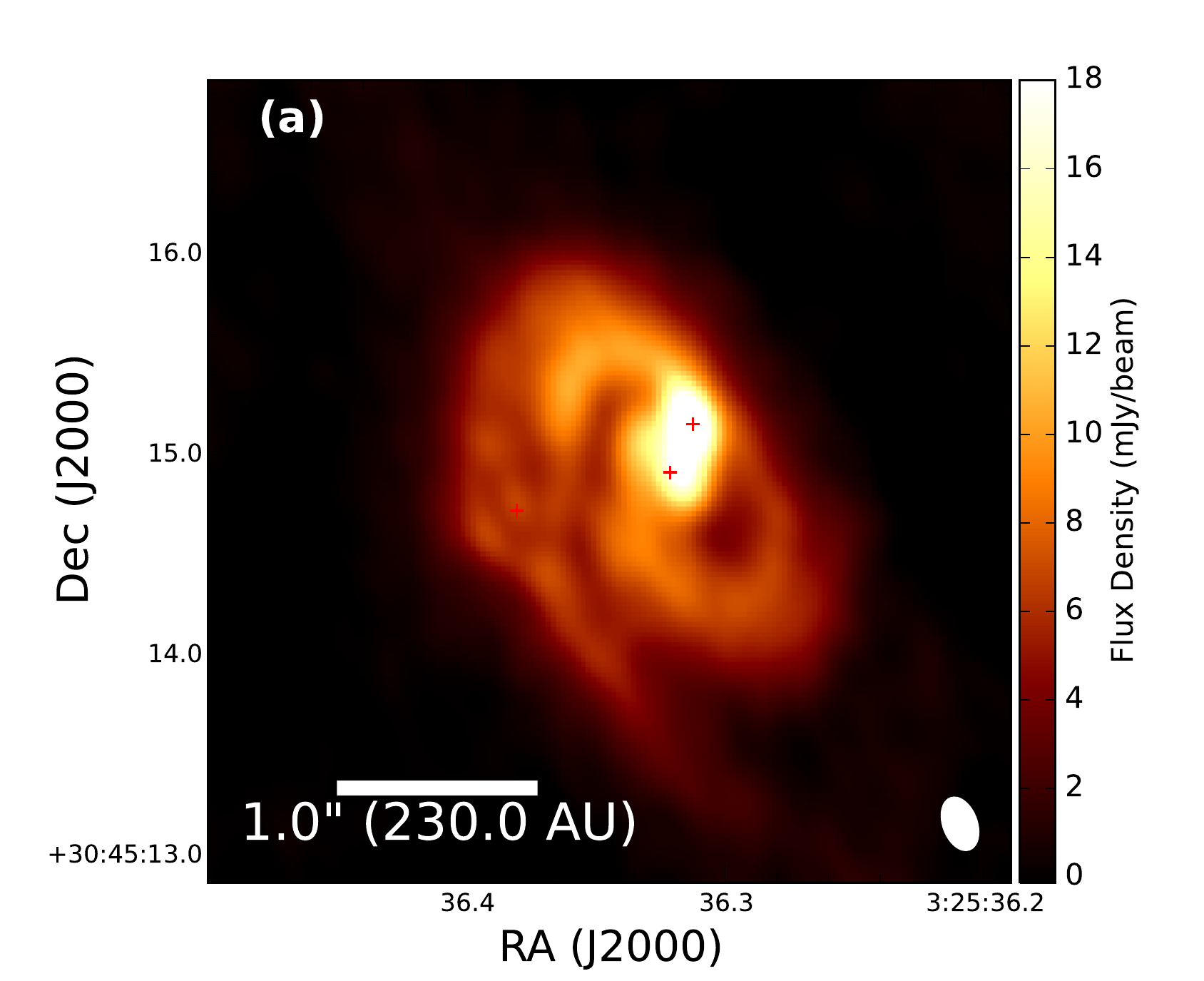}
\includegraphics[scale=0.7]{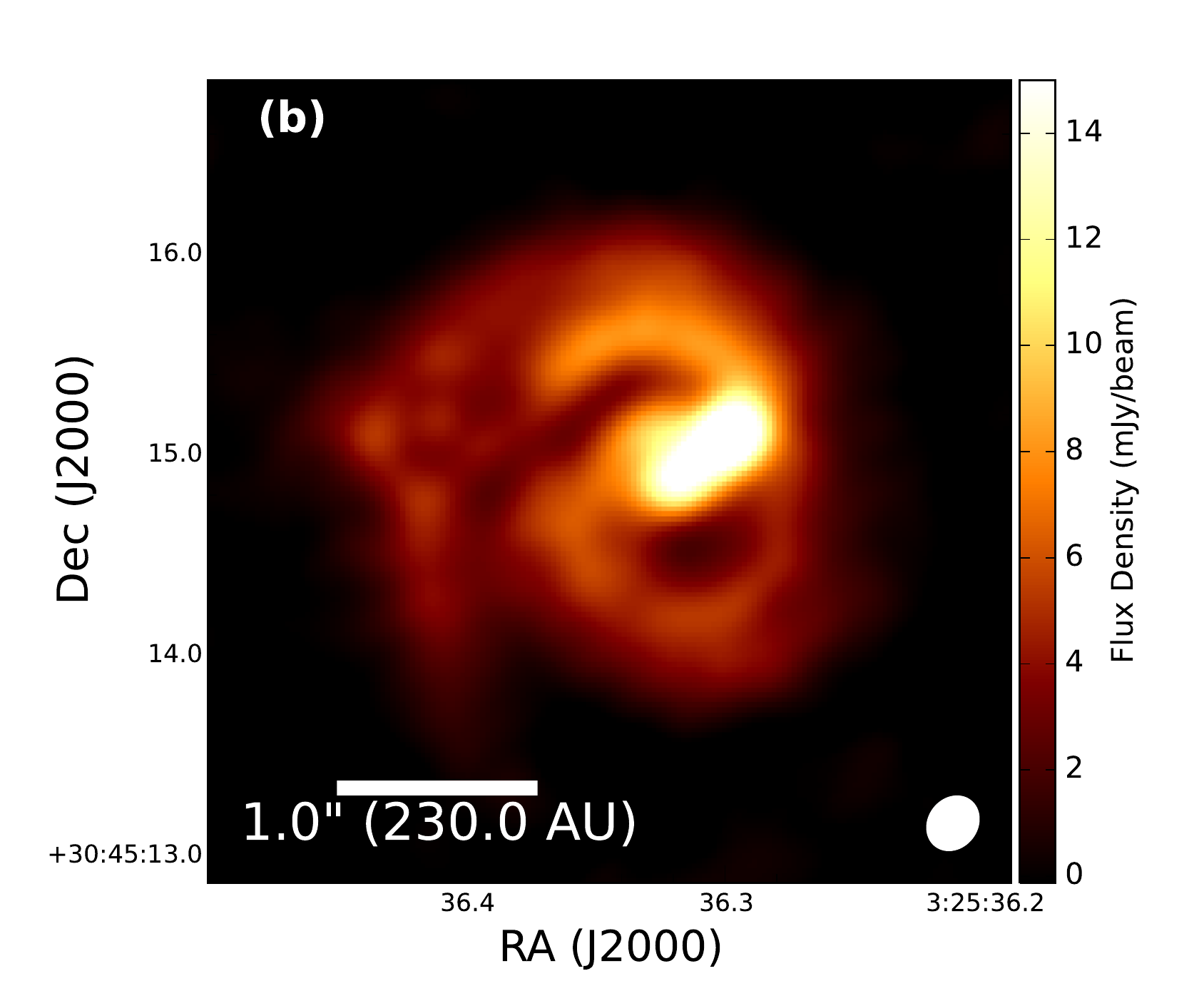}
\end{center}

\caption{ALMA 1.3~mm images with the brightest protostar (IRS3B-c)
removed. The image with IRS3B-c removed, as observed is shown in panel (a). Panel (b) shows the image 
deprojected (rotated and corrected for system inclination), also with IRS3B-c removed.
}
\end{figure}

\begin{figure}[!ht]
\begin{center}
\includegraphics[scale=1.0]{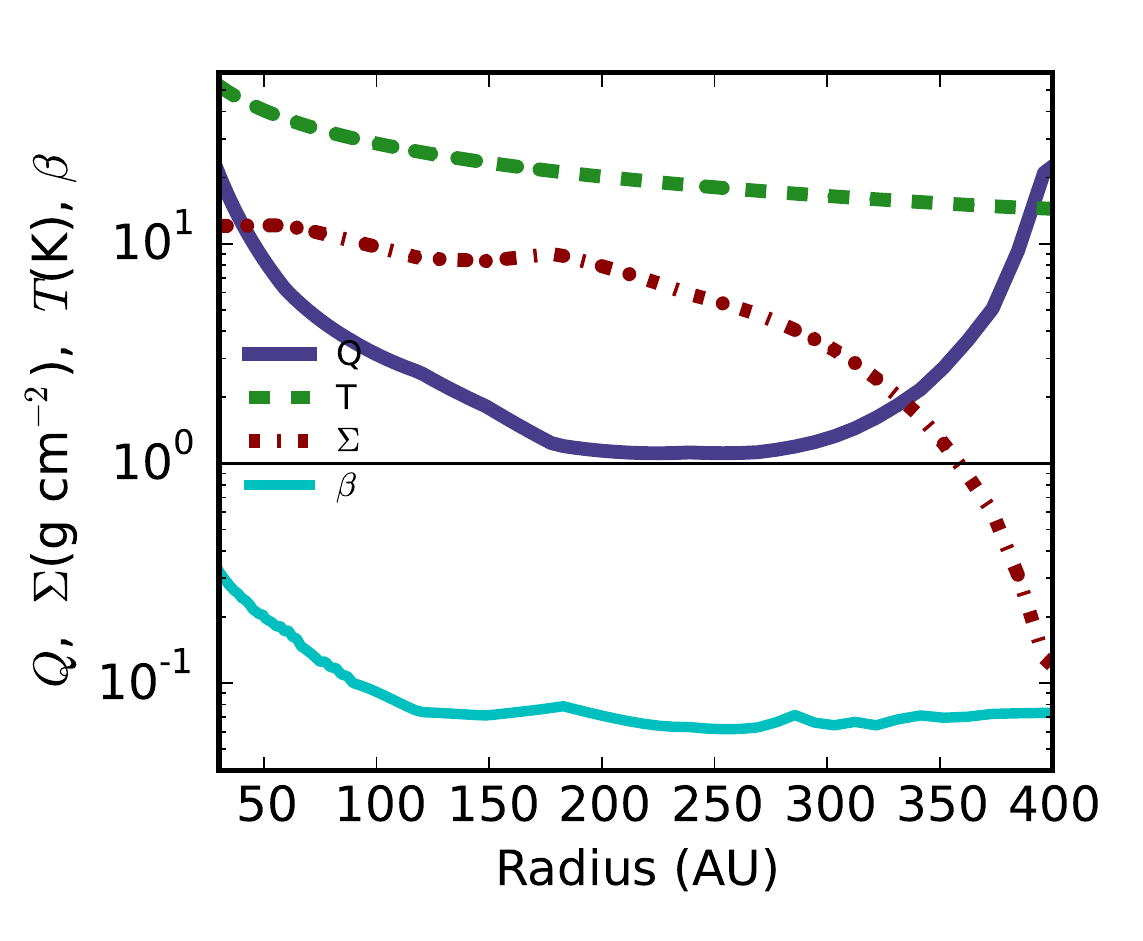}
\end{center}
\label{fig-selfcon}
\caption{Plot of observed disk surface density ($\Sigma$), Temperature (T), Toomre Q, and 
dimensionless cooling $\beta$. Temperature is plotted as the green line, Toomre Q (blue line) 
is calculated self-consistently from the inferred surface density profile (red) using the 
temperature dependent opacities\citep{semenov2003}. The cyan line indicates the 
dimensionless cooling rate, $\beta$. The black line demarcates unity, where the disk 
is expected to be gravitationally unstable. 
}
\end{figure}

\begin{figure}[!ht]
\begin{center}
\includegraphics[scale=0.4]{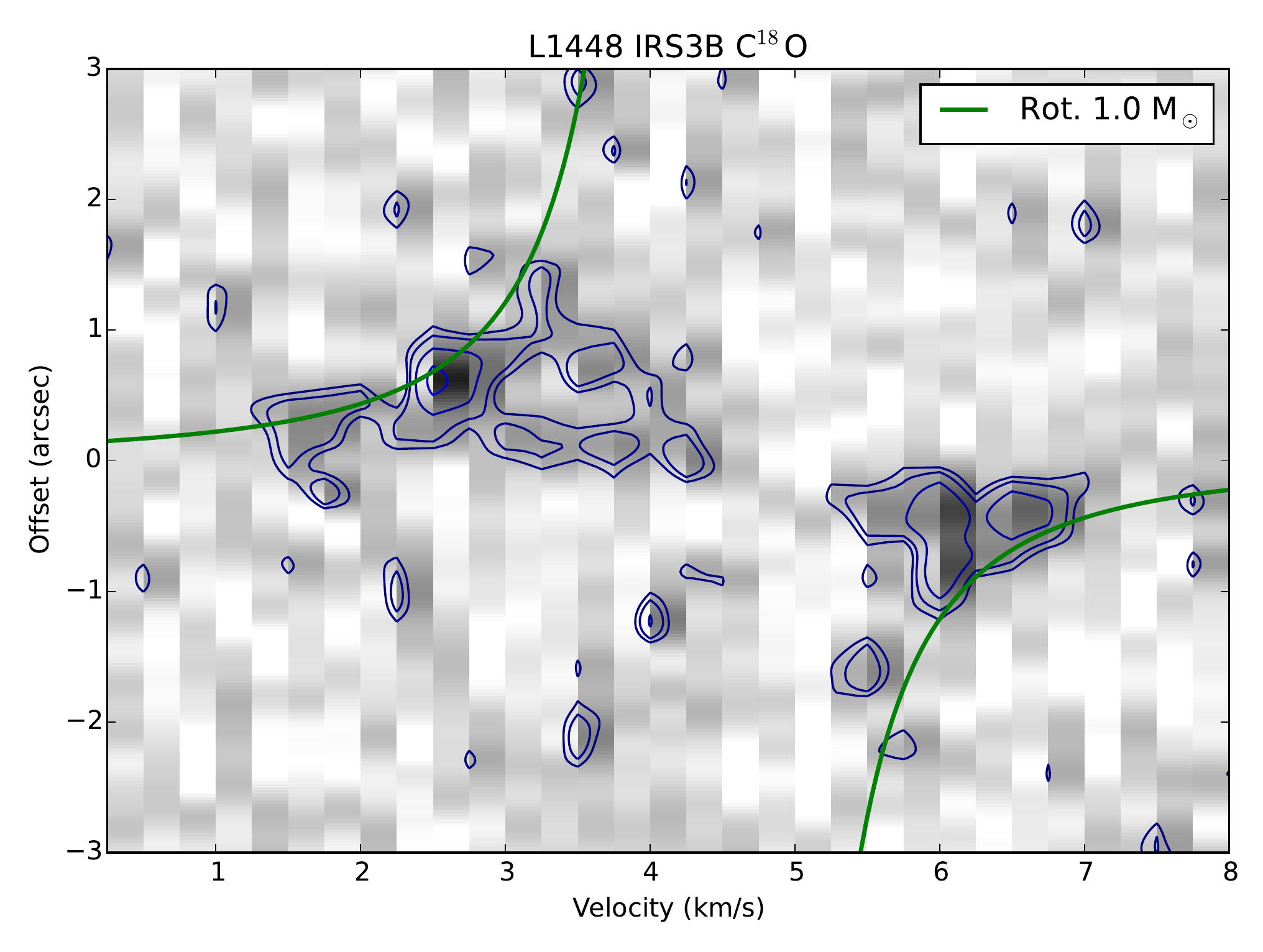}
\includegraphics[scale=0.4]{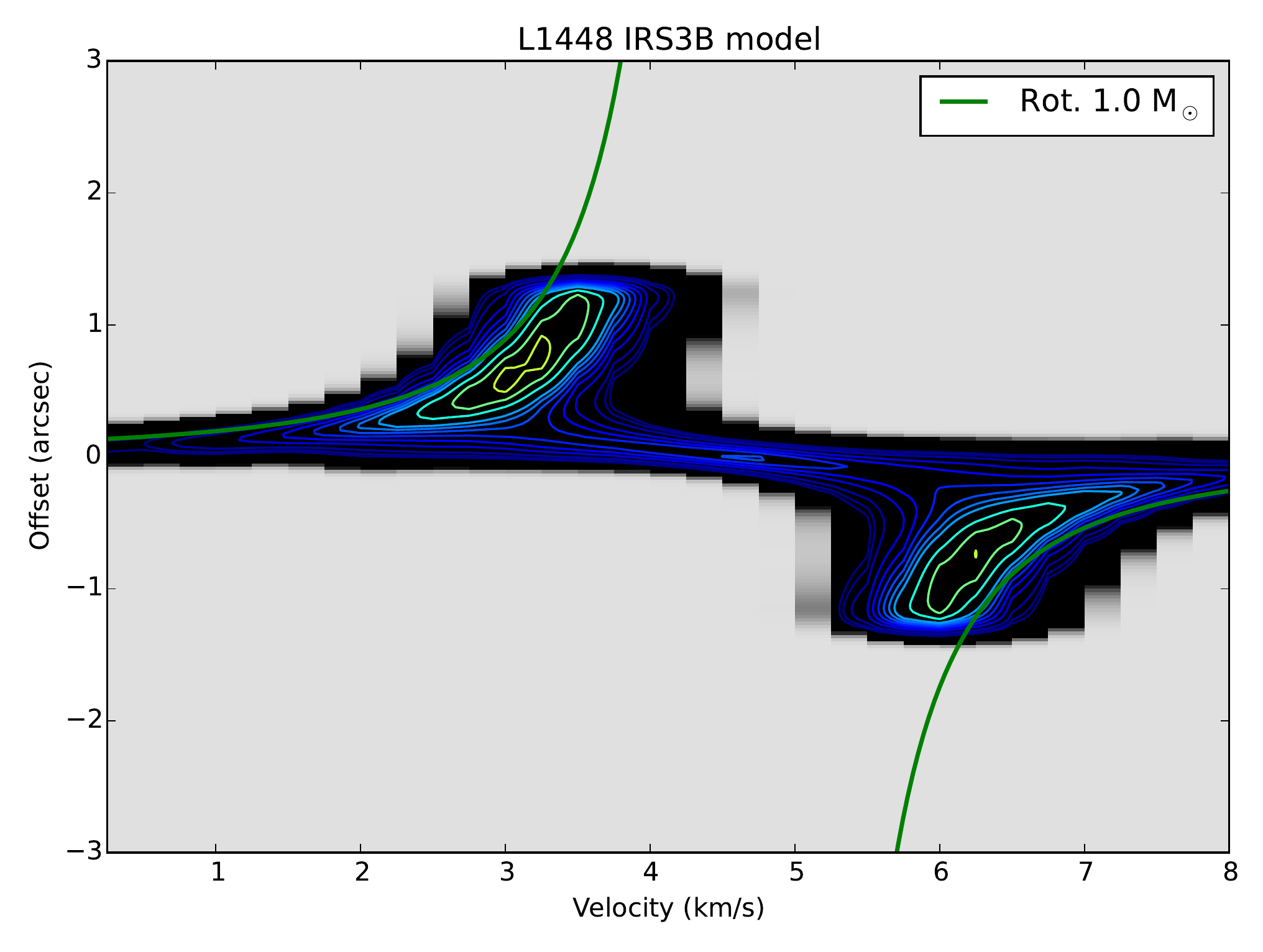}
\end{center}
\caption{Position-velocity diagrams of L1448 IRS3B and a model disk showing the rotation profile.
A position-velocity (PV) cut is taken along the major axis of the disk (analogous to a long-slit spectrum), 
across the position of IRS3B-a and IRS3B-b and shown in the top panel. 
The solid line drawn in the panel is a Keplerian rotation curve for
a 1.0~$M_{\sun}$ central protostar, assumed to be the combined mass of IRS3B-a/b. 
A PV diagram for a thin disk model is shown in the bottom panel, 
demonstrating that a model with the same inclination angle shows a consistent
PV diagram.
}
\end{figure}

\clearpage

\begin{small}

\bibliographystyle{naturemag}
\bibliography{ms}

\begin{thebibliography}{10}
\expandafter\ifx\csname url\endcsname\relax
  \def\url#1{\texttt{#1}}\fi
\expandafter\ifx\csname urlprefix\endcsname\relax\def\urlprefix{URL }\fi
\providecommand{\bibinfo}[2]{#2}
\providecommand{\eprint}[2][]{\url{#2}}

\bibitem{duchene2013}
\bibinfo{author}{{Duch{\^e}ne}, G.} \& \bibinfo{author}{{Kraus}, A.}
\newblock \bibinfo{title}{{Stellar Multiplicity}}.
\newblock \emph{\bibinfo{journal}{\araa}} \textbf{\bibinfo{volume}{51}},
  \bibinfo{pages}{269--310}
\newblock \eprint{ (\bibinfo{year}{2013})}.

\bibitem{reipurth2014}
\bibinfo{author}{{Reipurth}, B.} \emph{et~al.}
\newblock \bibinfo{title}{{Multiplicity in Early Stellar Evolution}}.
\newblock \emph{\bibinfo{journal}{Protostars and Planets VI}}
  \bibinfo{pages}{267--290}
\newblock \eprint{ (\bibinfo{year}{2014})}.

\bibitem{raghavan2010}
\bibinfo{author}{{Raghavan}, D.} \emph{et~al.}
\newblock \bibinfo{title}{{A Survey of Stellar Families: Multiplicity of
  Solar-type Stars}}.
\newblock \emph{\bibinfo{journal}{\apjs}} \textbf{\bibinfo{volume}{190}},
  \bibinfo{pages}{1--42}
\newblock \eprint{ (\bibinfo{year}{2010})}.

\bibitem{fisher2004}
\bibinfo{author}{{Fisher}, R.~T.}
\newblock \bibinfo{title}{{A Turbulent Interstellar Medium Origin of the Binary
  Period Distribution}}.
\newblock \emph{\bibinfo{journal}{\apj}} \textbf{\bibinfo{volume}{600}},
  \bibinfo{pages}{769--780}
\newblock \eprint{ (\bibinfo{year}{2004})}.

\bibitem{padoan2004}
\bibinfo{author}{{Padoan}, P.} \& \bibinfo{author}{{Nordlund}, {\AA}.}
\newblock \bibinfo{title}{{The ``Mysterious'' Origin of Brown Dwarfs}}.
\newblock \emph{\bibinfo{journal}{\apj}} \textbf{\bibinfo{volume}{617}},
  \bibinfo{pages}{559--564}
\newblock \eprint{ (\bibinfo{year}{2004})}.

\bibitem{adams1989}
\bibinfo{author}{{Adams}, F.~C.}, \bibinfo{author}{{Ruden}, S.~P.} \&
  \bibinfo{author}{{Shu}, F.~H.}
\newblock \bibinfo{title}{{Eccentric gravitational instabilities in nearly
  Keplerian disks}}.
\newblock \emph{\bibinfo{journal}{\apj}} \textbf{\bibinfo{volume}{347}},
  \bibinfo{pages}{959--976}
\newblock \eprint{ (\bibinfo{year}{1989})}.

\bibitem{bonnell1994b}
\bibinfo{author}{{Bonnell}, I.~A.} \& \bibinfo{author}{{Bate}, M.~R.}
\newblock \bibinfo{title}{{The Formation of Close Binary Systems}}.
\newblock \emph{\bibinfo{journal}{\mnras}} \textbf{\bibinfo{volume}{271}},
  \bibinfo{pages}{999--1004}
\newblock \eprint{ (\bibinfo{year}{1994})}.

\bibitem{pineda2015}
\bibinfo{author}{{Pineda}, J.~E.} \emph{et~al.}
\newblock \bibinfo{title}{{The formation of a quadruple star system with wide
  separation}}.
\newblock \emph{\bibinfo{journal}{\nat}} \textbf{\bibinfo{volume}{518}},
  \bibinfo{pages}{213--215}
\newblock \eprint{ (\bibinfo{year}{2015})}.

\bibitem{lee2016}
\bibinfo{author}{{Lee}, K.~I.} \emph{et~al.}
\newblock \bibinfo{title}{{Misalignment of Outflow Axes in the Proto-multiple
  Systems in Perseus}}.
\newblock \emph{\bibinfo{journal}{\apjl}} \textbf{\bibinfo{volume}{820}},
  \bibinfo{pages}{L2}
\newblock \eprint{ (\bibinfo{year}{2016})}.

\bibitem{connelley2008}
\bibinfo{author}{{Connelley}, M.~S.}, \bibinfo{author}{{Reipurth}, B.} \&
  \bibinfo{author}{{Tokunaga}, A.~T.}
\newblock \bibinfo{title}{{The Evolution of the Multiplicity of Embedded
  Protostars. II. Binary Separation Distribution and Analysis}}.
\newblock \emph{\bibinfo{journal}{\aj}} \textbf{\bibinfo{volume}{135}},
  \bibinfo{pages}{2526--2536}
\newblock \eprint{ (\bibinfo{year}{2008})}.

\bibitem{kraus2011}
\bibinfo{author}{{Kraus}, A.~L.}, \bibinfo{author}{{Ireland}, M.~J.},
  \bibinfo{author}{{Martinache}, F.} \& \bibinfo{author}{{Hillenbrand}, L.~A.}
\newblock \bibinfo{title}{{Mapping the Shores of the Brown Dwarf Desert. II.
  Multiple Star Formation in Taurus-Auriga}}.
\newblock \emph{\bibinfo{journal}{\apj}} \textbf{\bibinfo{volume}{731}},
  \bibinfo{pages}{8}
\newblock \eprint{ (\bibinfo{year}{2011})}.

\bibitem{takakuwa2012}
\bibinfo{author}{{Takakuwa}, S.} \emph{et~al.}
\newblock \bibinfo{title}{{A Keplerian Circumbinary Disk around the
  Protostellar System L1551 NE}}.
\newblock \emph{\bibinfo{journal}{\apj}} \textbf{\bibinfo{volume}{754}},
  \bibinfo{pages}{52}
\newblock \eprint{ (\bibinfo{year}{2012})}.

\bibitem{tobin2016}
\bibinfo{author}{{Tobin}, J.~J.} \emph{et~al.}
\newblock \bibinfo{title}{{The VLA Nascent Disk and Multiplicity Survey of
  Perseus Protostars (VANDAM). II. Multiplicity of Protostars in the Perseus
  Molecular Cloud}}.
\newblock \emph{\bibinfo{journal}{\apj}} \textbf{\bibinfo{volume}{818}},
  \bibinfo{pages}{73}
\newblock \eprint{ (\bibinfo{year}{2016})}.

\bibitem{lee2015}
\bibinfo{author}{{Lee}, K.~I.} \emph{et~al.}
\newblock \bibinfo{title}{{Mass Assembly of Stellar Systems and Their Evolution
  with the SMA (MASSES). Multiplicity and the Physical Environment in L1448N}}.
\newblock \emph{\bibinfo{journal}{\apj}} \textbf{\bibinfo{volume}{814}},
  \bibinfo{pages}{114}
\newblock \eprint{ (\bibinfo{year}{2015})}.

\bibitem{hirota2011}
\bibinfo{author}{{Hirota}, T.} \emph{et~al.}
\newblock \bibinfo{title}{{Astrometry of H$_{2}$O Masers in Nearby Star-Forming
  Regions with VERA. IV. L 1448 C}}.
\newblock \emph{\bibinfo{journal}{\pasj}} \textbf{\bibinfo{volume}{63}},
  \bibinfo{pages}{1}
\newblock \eprint{ (\bibinfo{year}{2011})}.

\bibitem{sadavoy2014}
\bibinfo{author}{{Sadavoy}, S.~I.} \emph{et~al.}
\newblock \bibinfo{title}{{Class 0 Protostars in the Perseus Molecular Cloud: A
  Correlation Between the Youngest Protostars and the Dense Gas Distribution}}.
\newblock \emph{\bibinfo{journal}{\apjl}} \textbf{\bibinfo{volume}{787}},
  \bibinfo{pages}{L18}
\newblock \eprint{ (\bibinfo{year}{2014})}.

\bibitem{andre1993}
\bibinfo{author}{{Andr\'e}, P.}, \bibinfo{author}{{Ward-Thompson}, D.} \&
  \bibinfo{author}{{Barsony}, M.}
\newblock \bibinfo{title}{{Submillimeter continuum observations of Rho Ophiuchi
  A - The candidate protostar VLA 1623 and prestellar clumps}}.
\newblock \emph{\bibinfo{journal}{\apj}} \textbf{\bibinfo{volume}{406}},
  \bibinfo{pages}{122--141}
\newblock \eprint{ (\bibinfo{year}{1993})}.

\bibitem{anglada1998}
\bibinfo{author}{{Anglada}, G.} \emph{et~al.}
\newblock \bibinfo{title}{{Spectral Indices of Centimeter Continuum Sources in
  Star-forming Regions: Implications on the Nature of the Outflow Exciting
  Sources}}.
\newblock \emph{\bibinfo{journal}{\aj}} \textbf{\bibinfo{volume}{116}},
  \bibinfo{pages}{2953--2964}
\newblock \eprint{ (\bibinfo{year}{1998})}.

\bibitem{toomre1964}
\bibinfo{author}{{Toomre}, A.}
\newblock \bibinfo{title}{{On the gravitational stability of a disk of stars}}.
\newblock \emph{\bibinfo{journal}{\apj}} \textbf{\bibinfo{volume}{139}},
  \bibinfo{pages}{1217--1238}
\newblock \eprint{ (\bibinfo{year}{1964})}.

\bibitem{kratter2016}
\bibinfo{author}{{Kratter}, K.~M.} \& \bibinfo{author}{{Lodato}, G.}
\newblock \bibinfo{title}{{Gravitational Instabilities in Circumstellar
  Disks}}.
\newblock \emph{\bibinfo{journal}{ArXiv e-prints}}
\newblock \eprint{ (\bibinfo{year}{2016})}.

\bibitem{kratter2010}
\bibinfo{author}{{Kratter}, K.~M.}, \bibinfo{author}{{Matzner}, C.~D.},
  \bibinfo{author}{{Krumholz}, M.~R.} \& \bibinfo{author}{{Klein}, R.~I.}
\newblock \bibinfo{title}{{On the Role of Disks in the Formation of Stellar
  Systems: A Numerical Parameter Study of Rapid Accretion}}.
\newblock \emph{\bibinfo{journal}{\apj}} \textbf{\bibinfo{volume}{708}},
  \bibinfo{pages}{1585--1597}
\newblock \eprint{ (\bibinfo{year}{2010})}.

\bibitem{bate2012}
\bibinfo{author}{{Bate}, M.~R.}
\newblock \bibinfo{title}{{Stellar, brown dwarf and multiple star properties
  from a radiation hydrodynamical simulation of star cluster formation}}.
\newblock \emph{\bibinfo{journal}{\mnras}} \textbf{\bibinfo{volume}{419}},
  \bibinfo{pages}{3115--3146}
\newblock \eprint{ (\bibinfo{year}{2012})}.

\bibitem{murillo2016}
\bibinfo{author}{{Murillo}, N.~M.}, \bibinfo{author}{{van Dishoeck}, E.~F.},
  \bibinfo{author}{{Tobin}, J.~J.} \& \bibinfo{author}{{Fedele}, D.}
\newblock \bibinfo{title}{{Do siblings always form and evolve simultaneously?
  Testing the coevality of multiple protostellar systems through SEDs}}.
\newblock \emph{\bibinfo{journal}{\aap}} \textbf{\bibinfo{volume}{592}},
  \bibinfo{pages}{A56}
\newblock \eprint{ (\bibinfo{year}{2016})}.

\bibitem{zhu2012}
\bibinfo{author}{{Zhu}, Z.}, \bibinfo{author}{{Hartmann}, L.},
  \bibinfo{author}{{Nelson}, R.~P.} \& \bibinfo{author}{{Gammie}, C.~F.}
\newblock \bibinfo{title}{{Challenges in Forming Planets by Gravitational
  Instability: Disk Irradiation and Clump Migration, Accretion, and Tidal
  Destruction}}.
\newblock \emph{\bibinfo{journal}{\apj}} \textbf{\bibinfo{volume}{746}},
  \bibinfo{pages}{110}
\newblock \eprint{ (\bibinfo{year}{2012})}.

\bibitem{Valtonen2008}
\bibinfo{author}{{Valtonen}, M.}, \bibinfo{author}{{Myll{\"a}ri}, A.},
  \bibinfo{author}{{Orlov}, V.} \& \bibinfo{author}{{Rubinov}, A.}
\newblock \bibinfo{title}{{The Problem of Three Stars: Stability Limit}}.
\newblock In \bibinfo{editor}{{Vesperini}, E.}, \bibinfo{editor}{{Giersz}, M.}
  \& \bibinfo{editor}{{Sills}, A.} (eds.) \emph{\bibinfo{booktitle}{Dynamical
  Evolution of Dense Stellar Systems}}, vol. \bibinfo{volume}{246} of
  \emph{\bibinfo{series}{IAU Symposium}}, \bibinfo{pages}{209--217}
\newblock \eprint{ (\bibinfo{year}{2008})}.

\bibitem{kiseleva1994}
\bibinfo{author}{{Kiseleva}, L.~G.}, \bibinfo{author}{{Eggleton}, P.~P.} \&
  \bibinfo{author}{{Orlov}, V.~V.}
\newblock \bibinfo{title}{{Instability of Close Triple Systems with Coplanar
  Initial Doubly Circular Motion}}.
\newblock \emph{\bibinfo{journal}{\mnras}} \textbf{\bibinfo{volume}{270}},
  \bibinfo{pages}{936}
\newblock \eprint{ (\bibinfo{year}{1994})}.

\bibitem{tokovinin2011}
\bibinfo{author}{{Tokovinin}, A.}
\newblock \bibinfo{title}{{Low-mass Visual Companions to Nearby G-dwarfs}}.
\newblock \emph{\bibinfo{journal}{\aj}} \textbf{\bibinfo{volume}{141}},
  \bibinfo{pages}{52}
\newblock \eprint{ (\bibinfo{year}{2011})}.

\bibitem{dipierro2014}
\bibinfo{author}{{Dipierro}, G.}, \bibinfo{author}{{Lodato}, G.},
  \bibinfo{author}{{Testi}, L.} \& \bibinfo{author}{{de Gregorio Monsalvo}, I.}
\newblock \bibinfo{title}{{How to detect the signatures of self-gravitating
  circumstellar discs with the Atacama Large Millimeter/sub-millimeter Array}}.
\newblock \emph{\bibinfo{journal}{\mnras}} \textbf{\bibinfo{volume}{444}},
  \bibinfo{pages}{1919--1929}
\newblock \eprint{ (\bibinfo{year}{2014})}.

\bibitem{grady2013}
\bibinfo{author}{{Grady}, C.~A.} \emph{et~al.}
\newblock \bibinfo{title}{{Spiral Arms in the Asymmetrically Illuminated Disk
  of MWC 758 and Constraints on Giant Planets}}.
\newblock \emph{\bibinfo{journal}{\apj}} \textbf{\bibinfo{volume}{762}},
  \bibinfo{pages}{48}
\newblock \eprint{ (\bibinfo{year}{2013})}.

\bibitem{dong2016}
\bibinfo{author}{{Dong}, R.}, \bibinfo{author}{{Vorobyov}, E.},
  \bibinfo{author}{{Pavlyuchenkov}, Y.}, \bibinfo{author}{{Chiang}, E.} \&
  \bibinfo{author}{{Liu}, H.~B.}
\newblock \bibinfo{title}{{Signatures of Gravitational Instability in Resolved
  Images of Protostellar Disks}}.
\newblock \emph{\bibinfo{journal}{\apj}} \textbf{\bibinfo{volume}{823}},
  \bibinfo{pages}{141}
\newblock \eprint{ (\bibinfo{year}{2016})}.
\clearpage
\section{References}
\bibitem{mcmullin2007}
\bibinfo{author}{{McMullin}, J.~P.}, \bibinfo{author}{{Waters}, B.},
  \bibinfo{author}{{Schiebel}, D.}, \bibinfo{author}{{Young}, W.} \&
  \bibinfo{author}{{Golap}, K.}
\newblock \bibinfo{title}{{CASA Architecture and Applications}}.
\newblock In \bibinfo{editor}{{Shaw}, R.~A.}, \bibinfo{editor}{{Hill}, F.} \&
  \bibinfo{editor}{{Bell}, D.~J.} (eds.) \emph{\bibinfo{booktitle}{Astronomical
  Data Analysis Software and Systems XVI}}, vol. \bibinfo{volume}{376} of
  \emph{\bibinfo{series}{Astronomical Society of the Pacific Conference
  Series}}, \bibinfo{pages}{127}
\newblock \eprint{ (\bibinfo{year}{2007})}.

\bibitem{looney2000}
\bibinfo{author}{{Looney}, L.~W.}, \bibinfo{author}{{Mundy}, L.~G.} \&
  \bibinfo{author}{{Welch}, W.~J.}
\newblock \bibinfo{title}{{Unveiling the Circumstellar Envelope and Disk: A
  Subarcsecond Survey of Circumstellar Structures}}.
\newblock \emph{\bibinfo{journal}{\apj}} \textbf{\bibinfo{volume}{529}},
  \bibinfo{pages}{477--498}
\newblock \eprint{ (\bibinfo{year}{2000})}.

\bibitem{sakai2014b}
\bibinfo{author}{{Sakai}, N.} \emph{et~al.}
\newblock \bibinfo{title}{{A Chemical View of Protostellar-disk Formation in
  L1527}}.
\newblock \emph{\bibinfo{journal}{\apjl}} \textbf{\bibinfo{volume}{791}},
  \bibinfo{pages}{L38}
\newblock \eprint{ (\bibinfo{year}{2014})}.

\bibitem{oya2014}
\bibinfo{author}{{Oya}, Y.} \emph{et~al.}
\newblock \bibinfo{title}{{A Substellar-mass Protostar and its Outflow of IRAS
  15398-3359 Revealed by Subarcsecond-resolution Observations of H$_{2}$CO and
  CCH}}.
\newblock \emph{\bibinfo{journal}{\apj}} \textbf{\bibinfo{volume}{795}},
  \bibinfo{pages}{152}
\newblock \eprint{ (\bibinfo{year}{2014})}.

\bibitem{wakelam2005}
\bibinfo{author}{{Wakelam}, V.} \emph{et~al.}
\newblock \bibinfo{title}{{Sulphur chemistry and molecular shocks: The case of
  NGC 1333-IRAS 2}}.
\newblock \emph{\bibinfo{journal}{\aap}} \textbf{\bibinfo{volume}{437}},
  \bibinfo{pages}{149--158}
\newblock \eprint{ (\bibinfo{year}{2005})}.

\bibitem{sakai2014a}
\bibinfo{author}{{Sakai}, N.} \emph{et~al.}
\newblock \bibinfo{title}{{Change in the chemical composition of infalling gas
  forming a disk around a protostar}}.
\newblock \emph{\bibinfo{journal}{\nat}} \textbf{\bibinfo{volume}{507}},
  \bibinfo{pages}{78--80}
\newblock \eprint{ (\bibinfo{year}{2014})}.

\bibitem{oya2016}
\bibinfo{author}{{Oya}, Y.} \emph{et~al.}
\newblock \bibinfo{title}{{Infalling-Rotating Motion and Associated Chemical
  Change in the Envelope of IRAS 16293-2422 Source A Studied with ALMA}}.
\newblock \emph{\bibinfo{journal}{\apj}} \textbf{\bibinfo{volume}{824}},
  \bibinfo{pages}{88}
\newblock \eprint{ (\bibinfo{year}{2016})}.

\bibitem{tobin2012}
\bibinfo{author}{{Tobin}, J.~J.} \emph{et~al.}
\newblock \bibinfo{title}{{A \~{}0.2-solar-mass protostar with a Keplerian disk
  in the very young L1527 IRS system}}.
\newblock \emph{\bibinfo{journal}{\nat}} \textbf{\bibinfo{volume}{492}},
  \bibinfo{pages}{83--85}
\newblock \eprint{ (\bibinfo{year}{2012})}.

\bibitem{harsono2014}
\bibinfo{author}{{Harsono}, D.} \emph{et~al.}
\newblock \bibinfo{title}{{Rotationally-supported disks around Class I sources
  in Taurus: disk formation constraints}}.
\newblock \emph{\bibinfo{journal}{\aap}} \textbf{\bibinfo{volume}{562}},
  \bibinfo{pages}{A77}
\newblock \eprint{ (\bibinfo{year}{2014})}.

\bibitem{jorgensen2009}
\bibinfo{author}{{J{\o}rgensen}, J.~K.} \emph{et~al.}
\newblock \bibinfo{title}{{PROSAC: a submillimeter array survey of low-mass
  protostars. II. The mass evolution of envelopes, disks, and stars from the
  Class 0 through I stages}}.
\newblock \emph{\bibinfo{journal}{\aap}} \textbf{\bibinfo{volume}{507}},
  \bibinfo{pages}{861--879}
\newblock \eprint{ (\bibinfo{year}{2009})}.

\bibitem{tobin2013}
\bibinfo{author}{{Tobin}, J.~J.} \emph{et~al.}
\newblock \bibinfo{title}{{Modeling the Resolved Disk around the Class 0
  Protostar L1527}}.
\newblock \emph{\bibinfo{journal}{\apj}} \textbf{\bibinfo{volume}{771}},
  \bibinfo{pages}{48}
\newblock \eprint{ (\bibinfo{year}{2013})}.

\bibitem{testi2014}
\bibinfo{author}{{Testi}, L.} \emph{et~al.}
\newblock \bibinfo{title}{{Dust Evolution in Protoplanetary Disks}}.
\newblock \emph{\bibinfo{journal}{Protostars and Planets VI}}
  \bibinfo{pages}{339--361}
\newblock \eprint{ (\bibinfo{year}{2014})}.

\bibitem{ossenkopf1994}
\bibinfo{author}{{Ossenkopf}, V.} \& \bibinfo{author}{{Henning}, T.}
\newblock \bibinfo{title}{{Dust opacities for protostellar cores}}.
\newblock \emph{\bibinfo{journal}{\aap}} \textbf{\bibinfo{volume}{291}},
  \bibinfo{pages}{943--959}
\newblock \eprint{ (\bibinfo{year}{1994})}.

\bibitem{andrews2010}
\bibinfo{author}{{Andrews}, S.~M.}, \bibinfo{author}{{Wilner}, D.~J.},
  \bibinfo{author}{{Hughes}, A.~M.}, \bibinfo{author}{{Qi}, C.} \&
  \bibinfo{author}{{Dullemond}, C.~P.}
\newblock \bibinfo{title}{{Protoplanetary Disk Structures in Ophiuchus. II.
  Extension to Fainter Sources}}.
\newblock \emph{\bibinfo{journal}{\apj}} \textbf{\bibinfo{volume}{723}},
  \bibinfo{pages}{1241--1254}
\newblock \eprint{ (\bibinfo{year}{2010})}.

\bibitem{ansdell2016}
\bibinfo{author}{{Ansdell}, M.} \emph{et~al.}
\newblock \bibinfo{title}{{ALMA Survey of Lupus Protoplanetary Disks I: Dust
  and Gas Masses}}.
\newblock \emph{\bibinfo{journal}{ArXiv e-prints}}
\newblock \eprint{ (\bibinfo{year}{2016})}.

\bibitem{beckwith1990}
\bibinfo{author}{{Beckwith}, S.~V.~W.}, \bibinfo{author}{{Sargent}, A.~I.},
  \bibinfo{author}{{Chini}, R.~S.} \& \bibinfo{author}{{Guesten}, R.}
\newblock \bibinfo{title}{{A survey for circumstellar disks around young
  stellar objects}}.
\newblock \emph{\bibinfo{journal}{\aj}} \textbf{\bibinfo{volume}{99}},
  \bibinfo{pages}{924--945}
\newblock \eprint{ (\bibinfo{year}{1990})}.

\bibitem{bohlin1978}
\bibinfo{author}{{Bohlin}, R.~C.}, \bibinfo{author}{{Savage}, B.~D.} \&
  \bibinfo{author}{{Drake}, J.~F.}
\newblock \bibinfo{title}{{A survey of interstellar H I from L-alpha absorption
  measurements. II}}.
\newblock \emph{\bibinfo{journal}{\apj}} \textbf{\bibinfo{volume}{224}},
  \bibinfo{pages}{132--142}
\newblock \eprint{ (\bibinfo{year}{1978})}.

\bibitem{bergin2013}
\bibinfo{author}{{Bergin}, E.~A.} \emph{et~al.}
\newblock \bibinfo{title}{{An old disk still capable of forming a planetary
  system}}.
\newblock \emph{\bibinfo{journal}{\nat}} \textbf{\bibinfo{volume}{493}},
  \bibinfo{pages}{644--646}
\newblock \eprint{ (\bibinfo{year}{2013})}.

\bibitem{williams2014}
\bibinfo{author}{{Williams}, J.~P.} \& \bibinfo{author}{{Best}, W.~M.~J.}
\newblock \bibinfo{title}{{A Parametric Modeling Approach to Measuring the Gas
  Masses of Circumstellar Disks}}.
\newblock \emph{\bibinfo{journal}{\apj}} \textbf{\bibinfo{volume}{788}},
  \bibinfo{pages}{59}
\newblock \eprint{ (\bibinfo{year}{2014})}.

\bibitem{perez2012}
\bibinfo{author}{{P{\'e}rez}, L.~M.} \emph{et~al.}
\newblock \bibinfo{title}{{Constraints on the Radial Variation of Grain Growth
  in the AS 209 Circumstellar Disk}}.
\newblock \emph{\bibinfo{journal}{\apjl}} \textbf{\bibinfo{volume}{760}},
  \bibinfo{pages}{L17}
\newblock \eprint{ (\bibinfo{year}{2012})}.

\bibitem{persson2016}
\bibinfo{author}{{Persson}, M.~V.} \emph{et~al.}
\newblock \bibinfo{title}{{Constraining the physical structure of the inner few
  100 AU scales of deeply embedded low-mass protostars}}.
\newblock \emph{\bibinfo{journal}{\aap}} \textbf{\bibinfo{volume}{590}},
  \bibinfo{pages}{A33}
\newblock \eprint{ (\bibinfo{year}{2016})}.

\bibitem{ohashi2014}
\bibinfo{author}{{Ohashi}, N.} \emph{et~al.}
\newblock \bibinfo{title}{{Formation of a Keplerian Disk in the Infalling
  Envelope around L1527 IRS: Transformation from Infalling Motions to Kepler
  Motions}}.
\newblock \emph{\bibinfo{journal}{\apj}} \textbf{\bibinfo{volume}{796}},
  \bibinfo{pages}{131}
\newblock \eprint{ (\bibinfo{year}{2014})}.

\bibitem{seifried2016}
\bibinfo{author}{{Seifried}, D.}, \bibinfo{author}{{S{\'a}nchez-Monge},
  {\'A}.}, \bibinfo{author}{{Walch}, S.} \& \bibinfo{author}{{Banerjee}, R.}
\newblock \bibinfo{title}{{Revealing the dynamics of Class 0 protostellar discs
  with ALMA}}.
\newblock \emph{\bibinfo{journal}{\mnras}} \textbf{\bibinfo{volume}{459}},
  \bibinfo{pages}{1892--1906}
\newblock \eprint{ (\bibinfo{year}{2016})}.

\bibitem{oya2015}
\bibinfo{author}{{Oya}, Y.} \emph{et~al.}
\newblock \bibinfo{title}{{Geometric and Kinematic Structure of the
  Outflow/Envelope System of L1527 Revealed by Subarcsecond-resolution
  Observation of CS}}.
\newblock \emph{\bibinfo{journal}{\apj}} \textbf{\bibinfo{volume}{812}},
  \bibinfo{pages}{59}
\newblock \eprint{ (\bibinfo{year}{2015})}.

\bibitem{maret2015}
\bibinfo{author}{Maret, S.}
\newblock \bibinfo{title}{{thindisk 1.0: Compute the line emission from a
  geometrically thin protoplanetary disk}} (\bibinfo{year}{2015}).
\newblock
\newblock \eprint{\urlprefix\url{http://dx.doi.org/10.5281/zenodo.13823}}.

\bibitem{rafi2005}
\bibinfo{author}{{Rafikov}, R.~R.}
\newblock \bibinfo{title}{{Can Giant Planets Form by Direct Gravitational
  Instability?}}
\newblock \emph{\bibinfo{journal}{\apjl}} \textbf{\bibinfo{volume}{621}},
  \bibinfo{pages}{L69--L72}
\newblock \eprint{ (\bibinfo{year}{2005})}.

\bibitem{gammie2001}
\bibinfo{author}{{Gammie}, C.~F.}
\newblock \bibinfo{title}{{Nonlinear Outcome of Gravitational Instability in
  Cooling, Gaseous Disks}}.
\newblock \emph{\bibinfo{journal}{\apj}} \textbf{\bibinfo{volume}{553}},
  \bibinfo{pages}{174--183}
\newblock \eprint{ (\bibinfo{year}{2001})}.

\bibitem{meru2011}
\bibinfo{author}{{Meru}, F.} \& \bibinfo{author}{{Bate}, M.~R.}
\newblock \bibinfo{title}{{Non-convergence of the critical cooling time-scale
  for fragmentation of self-gravitating discs}}.
\newblock \emph{\bibinfo{journal}{\mnras}} \textbf{\bibinfo{volume}{411}},
  \bibinfo{pages}{L1--L5}
\newblock \eprint{ (\bibinfo{year}{2011})}.

\bibitem{paardekooper2012}
\bibinfo{author}{{Paardekooper}, S.-J.}
\newblock \bibinfo{title}{{Numerical convergence in self-gravitating shearing
  sheet simulations and the stochastic nature of disc fragmentation}}.
\newblock \emph{\bibinfo{journal}{\mnras}} \textbf{\bibinfo{volume}{421}},
  \bibinfo{pages}{3286--3299}
\newblock \eprint{ (\bibinfo{year}{2012})}.

\bibitem{lodato2011}
\bibinfo{author}{{Lodato}, G.} \& \bibinfo{author}{{Clarke}, C.~J.}
\newblock \bibinfo{title}{{Resolution requirements for smoothed particle
  hydrodynamics simulations of self-gravitating accretion discs}}.
\newblock \emph{\bibinfo{journal}{\mnras}} \textbf{\bibinfo{volume}{413}},
  \bibinfo{pages}{2735--2740}
\newblock \eprint{ (\bibinfo{year}{2011})}.

\bibitem{baraffe2015}
\bibinfo{author}{{Baraffe}, I.}, \bibinfo{author}{{Homeier}, D.},
  \bibinfo{author}{{Allard}, F.} \& \bibinfo{author}{{Chabrier}, G.}
\newblock \bibinfo{title}{{New evolutionary models for pre-main sequence and
  main sequence low-mass stars down to the hydrogen-burning limit}}.
\newblock \emph{\bibinfo{journal}{\aap}} \textbf{\bibinfo{volume}{577}},
  \bibinfo{pages}{A42}
\newblock \eprint{ (\bibinfo{year}{2015})}.

\bibitem{kratter2008}
\bibinfo{author}{{Kratter}, K.~M.}, \bibinfo{author}{{Matzner}, C.~D.} \&
  \bibinfo{author}{{Krumholz}, M.~R.}
\newblock \bibinfo{title}{{Global Models for the Evolution of Embedded,
  Accreting Protostellar Disks}}.
\newblock \emph{\bibinfo{journal}{\apj}} \textbf{\bibinfo{volume}{681}},
  \bibinfo{pages}{375--390}
\newblock \eprint{ (\bibinfo{year}{2008})}.

\bibitem{tsc1984}
\bibinfo{author}{{Terebey}, S.}, \bibinfo{author}{{Shu}, F.~H.} \&
  \bibinfo{author}{{Cassen}, P.}
\newblock \bibinfo{title}{{The collapse of the cores of slowly rotating
  isothermal clouds}}.
\newblock \emph{\bibinfo{journal}{\apj}} \textbf{\bibinfo{volume}{286}},
  \bibinfo{pages}{529--551}
\newblock \eprint{ (\bibinfo{year}{1984})}.

\bibitem{matzner2005}
\bibinfo{author}{{Matzner}, C.~D.} \& \bibinfo{author}{{Levin}, Y.}
\newblock \bibinfo{title}{{Protostellar Disks: Formation, Fragmentation, and
  the Brown Dwarf Desert}}.
\newblock \emph{\bibinfo{journal}{\apj}} \textbf{\bibinfo{volume}{628}},
  \bibinfo{pages}{817--831}
\newblock \eprint{ (\bibinfo{year}{2005})}.

\bibitem{pollack1985}
\bibinfo{author}{{Pollack}, J.~B.}, \bibinfo{author}{{McKay}, C.~P.} \&
  \bibinfo{author}{{Christofferson}, B.~M.}
\newblock \bibinfo{title}{{A calculation of the Rosseland mean opacity of dust
  grains in primordial solar system nebulae}}.
\newblock \emph{\bibinfo{journal}{\icarus}} \textbf{\bibinfo{volume}{64}},
  \bibinfo{pages}{471--492}
\newblock \eprint{ (\bibinfo{year}{1985})}.

\bibitem{semenov2003}
\bibinfo{author}{{Semenov}, D.}, \bibinfo{author}{{Henning}, T.},
  \bibinfo{author}{{Helling}, C.}, \bibinfo{author}{{Ilgner}, M.} \&
  \bibinfo{author}{{Sedlmayr}, E.}
\newblock \bibinfo{title}{{Rosseland and Planck mean opacities for
  protoplanetary discs}}.
\newblock \emph{\bibinfo{journal}{\aap}} \textbf{\bibinfo{volume}{410}},
  \bibinfo{pages}{611--621}
\newblock \eprint{ (\bibinfo{year}{2003})}.

\bibitem{chiang1997}
\bibinfo{author}{{Chiang}, E.~I.} \& \bibinfo{author}{{Goldreich}, P.}
\newblock \bibinfo{title}{{Spectral Energy Distributions of T Tauri Stars with
  Passive Circumstellar Disks}}.
\newblock \emph{\bibinfo{journal}{\apj}} \textbf{\bibinfo{volume}{490}},
  \bibinfo{pages}{368--376}
\newblock \eprint{ (\bibinfo{year}{1997})}.

\bibitem{rice2005}
\bibinfo{author}{{Rice}, W.~K.~M.}, \bibinfo{author}{{Lodato}, G.} \&
  \bibinfo{author}{{Armitage}, P.~J.}
\newblock \bibinfo{title}{{Investigating fragmentation conditions in
  self-gravitating accretion discs}}.
\newblock \emph{\bibinfo{journal}{\mnras}} \textbf{\bibinfo{volume}{364}},
  \bibinfo{pages}{L56--L60}
\newblock \eprint{ (\bibinfo{year}{2005})}.

\bibitem{shu1977}
\bibinfo{author}{{Shu}, F.~H.}
\newblock \bibinfo{title}{{Self-similar collapse of isothermal spheres and star
  formation}}.
\newblock \emph{\bibinfo{journal}{\apj}} \textbf{\bibinfo{volume}{214}},
  \bibinfo{pages}{488--497}
\newblock \eprint{ (\bibinfo{year}{1977})}.

\bibitem{enoch2009}
\bibinfo{author}{{Enoch}, M.~L.}, \bibinfo{author}{{Evans}, N.~J.},
  \bibinfo{author}{{Sargent}, A.~I.} \& \bibinfo{author}{{Glenn}, J.}
\newblock \bibinfo{title}{{Properties of the Youngest Protostars in Perseus,
  Serpens, and Ophiuchus}}.
\newblock \emph{\bibinfo{journal}{\apj}} \textbf{\bibinfo{volume}{692}},
  \bibinfo{pages}{973--997}
\newblock \eprint{ (\bibinfo{year}{2009})}.

\bibitem{offner2011}
\bibinfo{author}{{Offner}, S.~S.~R.} \& \bibinfo{author}{{McKee}, C.~F.}
\newblock \bibinfo{title}{{The Protostellar Luminosity Function}}.
\newblock \emph{\bibinfo{journal}{\apj}} \textbf{\bibinfo{volume}{736}},
  \bibinfo{pages}{53}
\newblock \eprint{ (\bibinfo{year}{2011})}.

\bibitem{whitney2003a}
\bibinfo{author}{{Whitney}, B.~A.}, \bibinfo{author}{{Wood}, K.},
  \bibinfo{author}{{Bjorkman}, J.~E.} \& \bibinfo{author}{{Wolff}, M.~J.}
\newblock \bibinfo{title}{{Two-dimensional Radiative Transfer in Protostellar
  Envelopes. I. Effects of Geometry on Class I Sources}}.
\newblock \emph{\bibinfo{journal}{\apj}} \textbf{\bibinfo{volume}{591}},
  \bibinfo{pages}{1049--1063}
\newblock \eprint{ (\bibinfo{year}{2003})}.

\bibitem{boley2010}
\bibinfo{author}{{Boley}, A.~C.}, \bibinfo{author}{{Hayfield}, T.},
  \bibinfo{author}{{Mayer}, L.} \& \bibinfo{author}{{Durisen}, R.~H.}
\newblock \bibinfo{title}{{Clumps in the outer disk by disk instability: Why
  they are initially gas giants and the legacy of disruption}}.
\newblock \emph{\bibinfo{journal}{\icarus}} \textbf{\bibinfo{volume}{207}},
  \bibinfo{pages}{509--516}
\newblock \eprint{ (\bibinfo{year}{2010})}.

\end{thebibliography}

\end{small}

\end{document}